# Consistency Models with Global Operation Sequencing and their Composition (Extended Version)

**Alexey Gotsman[1] and Sebastian Burckhardt[2]**

1 **IMDEA Software Institute, Spain**
2 **Microsoft Research, USA**

## Abstract

Modern distributed systems often achieve availability and scalability by providing consistency guarantees about the data they manage weaker than linearizability. We consider a class of such consistency models that, despite this weakening, guarantee that clients eventually agree on a global sequence of operations, while seeing a subsequence of this final sequence at any given point of time. Examples of such models include the classical Total Store Order (TSO) and recently proposed dual TSO, Global Sequence Protocol (GSP) and Ordered Sequential Consistency.

We define a unified model, called *Global Sequence Consistency (GSC)*, that has the above models as its special cases, and investigate its key properties. First, we propose a condition under which multiple objects each satisfying GSC can be composed so that the whole set of objects satisfies GSC. Second, we prove an interesting relationship between special cases of GSC—GSP, TSO and dual TSO: we show that clients that do not communicate out-of-band cannot tell the difference between these models. To obtain these results, we propose a novel axiomatic specification of GSC and prove its equivalence to the operational definition of the model.

## 1 Introduction

Modern distributed systems often achieve availability and scalability by providing consistency guarantees about the data they manage weaker than the gold standard of linearizability [15]. In this paper we consider a class of such consistency models that, despite this weakening, guarantee *global operation sequencing*: clients eventually agree on a global sequence of operations, while seeing a subsequence of this final sequence at any given point of time. An implementation of a service providing such a model may consist of a single *server* and multiple *clients*, each maintaining a replica of the data managed by the service. Clients accept operations from end-users, evaluate them on their local (possibly stale) data replica and forward the operations to the server. The server arranges all received operations into a totally ordered log and forwards them to clients in the order determined by the log. The server log thus establishes the desired global sequence of operations.

Such consistency models arise in different domains. For instance, clients may correspond to mobile devices, cloud servers or processor cores; the role of the server may be played by an elected leader, a replicated state machine [25], a reliable total-order broadcast [11] or the memory subsystem in a multiprocessor architecture [27]. Various models differ in whether the propagation of operations from clients to the server and vice versa is asynchronous or synchronous. Thus, in the *Global Sequence Protocol (GSP)* model [10], the propagation is asynchronous in both directions, which allows clients to execute operations even if they get partitioned from the server [14]. This model is implemented in Microsoft's TouchDevelop system for mobile app programming, to support offline access [1], and in the Orleans actor framework [6], to support geo-replication [5]. In the *Total Store Order (TSO)* model [22, 23], implemented by SPARC and x86 multiprocessors, operation propagation from clients to the server is asynchronous, but the one from the server to clients is synchronous: clients *pull* all new operations from the server before evaluating each operation. Conversely, in the *dual TSO* model [2] operation propagation from the server to clients is asynchronous, but the one from the clients to the server is synchronous: clients *push* operations to the server immediately after they are executed. If we strengthen dual TSO by requiring that all update operations are propagated synchronously in both directions, we obtain *Ordered Sequential Consistency (OSC)* [21], which captures the semantics of coordination services such as ZooKeeper [17]. Finally, we obtain linearizability [15] when operation propagation is synchronous in both directions.

In this paper we study key properties of the consistency models from the above class. To this end, we consider a flexible model, called *Global Sequence Consistency (GSC)*, that has the above models as its special cases and obtain novel results about this model: a condition for safely composing multiple GSC services and a certain interesting relationship between the model's special cases. The GSC model is defined by the above client-server protocol where operation propagation is by default asynchronous, but operations may include two kinds of *fences*. The fences respectively force a client to pull all new operations from the server or push all outstanding local operations to the server (§3). Then we obtain various existing consistency models by systematically associating fences with operations as shown in Figure 1.

| Implicit fences | pull | push |
|---|---|---|
| GSP [10] | no | no |
| TSO [22, 23] | yes | no |
| dual TSO [2] | no | yes |
| OSC [21] | updates | yes |
| linearizability [15] | yes | yes |

**Figure 1** Specialising GSC.

Like sequential consistency [19], GSC is not *composable* (aka *local*) [15]: objects satisfying GSC may fail to provide this consistency guarantee when combined. This is a problem because application programmers often want to distribute objects among multiple services, e.g., to place them in geographical locations where they are most likely to be updated and thereby minimise latency [20]. Non-composability does not allow programmers to easily predict the behavior of such a system. This is a particular issue in the Orleans implementation of geo-replication [5], which guarantees GSP only for each individual object.

To address this problem, we propose a condition under which multiple objects each satisfying GSC can be composed so that the whole set of objects satisfies GSC (§5). Informally, the condition requires using fences according to the following discipline: when switching between different objects, a client has to push the operations done on the old object and pull operations on the new object. Our result ensures that in this case clients interacting with multiple GSC services implementing different objects will behave as though they are interacting with a single GSC service. This result holds even when clients can communicate out-of-band, without using the GSC services. As its special cases, we obtain novel conditions for composing TSO and dual TSO objects, as well as a recently proposed condition for OSC [20, 21].

We also prove an interesting relationship between special cases of GSC—GSP, TSO and dual TSO (§4): we show that clients that do not communicate out-of-band cannot tell the difference between them. In particular, this result implies that a program without out-of-band communication written assuming TSO operates correctly under much weaker, fully asynchronous GSP. This equivalence has been previously conjectured without proof [10]; the present paper confirms this conjecture. Assuming the absence of out-of-band communication is common for memory models, where clients are processors that do not communicate directly. However, this assumption is often not appropriate for distributed interactive applications, where clients can have external means of communication. In this setting, the above special cases of GSC are observably different.

Proving the above results about compositionality and equivalence is nontrivial due to the complexity of reasoning about the distributed protocol implementing GSC. Our main tool in tackling this complexity is an *axiomatic* specification of GSC, given in the style often used for consistency models in shared-memory [18] and distributed storage systems [8, 9] (§6). The specification represents service executions using several relations, declaratively describing how operations are processed by the GSC protocol; the consistency model is then defined by a set of axioms, constraining these relations. We prove that our axiomatic specification is equivalent to the operational one. A particular subtlety in formulating the axiomatic specification and proving this equivalence is the need for the specification to track the *real-time order* between operations, determining when one operation finishes before another one starts. This makes results established using the axiomatic specification applicable in the case when clients can communicate out-of-band [3, 12].

The axiomatic specification of GSC is instrumental in obtaining our results. A recurring challenge is to prove the existence of an execution that satisfies some conditions, e.g., is a composition of single-object executions in the proof of the compositionality criterion (§8). Constructing the desired execution is difficult to do directly on the operational model. Because of the wide-ranging effect of fences, such an execution cannot be obtained simply by local reordering of independent steps, as with simpler operational models. But via the axiomatic specification of GSC, we can solve this problem indirectly by formulating constraints on precedence of events in the execution as relations and then using algebraic techniques to prove that their union is acyclic, which guarantees that there exists an execution satisfying them. We hope that, in the future, the GSC model, with its two equivalent definitions, and our proof techniques will provide a solid foundation for obtaining further results about consistency models with global operation sequencing.

## 2 Preliminaries

We consider a distributed service managing a collection of *objects* $\mathsf{Obj} = \{x, y, \ldots\}$. A finite number of clients interact with the service by performing *operations* on the objects, which are ranged over by $op$ and come from a set $\mathsf{Op}$. Parameters of operations, if any, are part of the operation name. For uniformity, we assume that all objects admit the same set of operations and that each operation returns one value from a set $\mathsf{Val}$; we can use a special member of $\mathsf{Val}$ to model operations that return no value. The sequential semantics of operations is defined by a function $\mathsf{eval} : \mathsf{Op}^* \times \mathsf{Op} \to \mathsf{Val}$ that determines the return value of an operation on an object given the sequence of operations previously executed on this object.

The consistency model provided by the service defines the set of all possible interactions between the service and its clients. We now introduce a structure that records such interactions in a single computation, called a *history*. In it we denote client-service interactions using *events*, which are ranged over by $e, f, g$ and come from an infinite countable set $\mathsf{Event}$. Events have unique identifiers from a set $\mathsf{Id}$. An event is of the form $e = (\iota, x, op, a, fen)$, where $\iota \in \mathsf{Id}$ is the event identifier, $x \in \mathsf{Obj}$ is the object on which the event occurs, $op \in \mathsf{Op}$ is the operation done, $a \in \mathsf{Val}$ is its return value, and $fen \subseteq \{\mathsf{push}, \mathsf{pull}\}$ gives the *fences* requested by the client. We use $\mathsf{obj}(e)$, $\mathsf{oper}(e)$, $\mathsf{rval}(e)$, $\mathsf{fences}(e)$ to select event components.

**State for each client** $c$:
```
  known_c ∈ (Id × Op)*
  unacked_c ∈ (Id × Op)*
  pending_c ∈ (Id × Op)*
exec(c, op, fen):
  if (pull ∈ fen)
    while (known_c ≠ server_log) pull(c)
  result :=
    eval(stripIds(known_c · unacked_c · pending_c), op)
  pending_c := pending_c · (uniqueId(), op)
  if (push ∈ fen)
    while (pending_c ≠ []) push(c)
  return result
```

**Server state:**
```
  server_log ∈ (Id × Op)*
push(c):
  if (pending_c = (id, op) · remaining_c)
    server_log := server_log · (id, op)
    unacked_c := unacked_c · (id, op)
    pending_c := remaining_c
pull(c):
  if (server_log = known_c · (id, op) · _)
    known_c := known_c · (id, op)
    if (unacked_c = (id, op) · remaining_c)
      unacked_c := remaining_c
```

**Figure 2** The pseudocode of the protocol defining the GSC model. We denote sequence concatenation by $\cdot$, an empty sequence by $[\,]$ and an irrelevant expression by $_$.

We use the following kinds of relations. A relation is a *strict partial order* if it is transitive and irreflexive. It is a *total order* if it additionally relates every two distinct elements one way or another. A relation is *prefix-finite* if each element is reachable along directed paths from at most finitely many others. A strict partial order $R$ is an *interval order* if

$$\forall e_1, e_2, f_1, f_2.\, (e_1 \xrightarrow{R} e_2 \wedge f_1 \xrightarrow{R} f_2) \implies (e_1 \xrightarrow{R} f_2 \vee f_1 \xrightarrow{R} e_2).$$

Intuitively, an interval order $R$ is consistent with an interpretation of events as segments of time during which the corresponding operations executed, with $R$ ordering $e$ before $f$ if $e$ finishes before $f$ starts [13]. For example, the real-time order considered in linearizability [15] is an interval order.

A *history* is a triple $\mathcal{H} = (E, \mathsf{so}, \mathsf{rt})$, where: $E \subseteq \mathsf{Event}$; *session order* $\mathsf{so} \subseteq E \times E$ is a union of prefix-finite total orders over a finite number of disjoint subsets of $E$ (each corresponding to operations by the same client); and *real-time order* $\mathsf{rt} \subseteq E \times E$ is a prefix-finite interval order such that $\mathsf{so} \subseteq \mathsf{rt}$ and $\forall e \in E.\, |\{f \in E \mid \neg(e \xrightarrow{\mathsf{rt}} f)\}| < \infty$.

The set $E$ defines all operations invoked by clients in a single computation and can be infinite. The session order arranges operations by the same client in the order in which they were executed. The real-time order $e \xrightarrow{\mathsf{rt}} f$ tells us that the operation of $e$ finished before the one of $f$ started (the last restriction on $\mathsf{rt}$ ensures that every operation finishes). Tracking this relationship is important because it allows the client who executed the operation of $e$ to communicate its return value to the client executing $f$ out-of-band, without using the service; the return value of $e$ can then influence the operation executed by $f$ [3, 12]. We denote components of histories and similar structures as in $E_{\mathcal{H}}$ and $\mathsf{so}_{\mathcal{H}}$. A consistency model is defined by a set of histories.

## 3 Operational Specification

We define Global Sequence Consistency using the idealised protocol in Figure 2, which is a generalisation of the Global Sequence Protocol (GSP) [10]. It assumes a single *server* and a finite number of *clients*. The server state is represented by a log *server_log* of operations received from clients, tagged with unique identifiers from $\mathsf{Id}$. The state of each client $c$ includes three logs: $known_c$ is the prefix of *server_log* that $c$ knows about; $pending_c$ is the log of operations by $c$ that have not yet been pushed to the server; and $unacked_c$ is the log of operations by $c$ that have been pushed to the server, but $known_c$ has not yet advanced enough to incorporate them.

The communication between the server and each client $c$ is modeled by transitions $\texttt{push}(c)$ and $\texttt{pull}(c)$ that can fire nondeterministically at any time when the client is not executing an operation and atomically modify the client and the server state (implementations may refine this using asynchronous communication channels as in [10]). The $\texttt{push}(c)$ function models how the server processes the next operation by client $c$: it appends the oldest record in $pending_c$ to $server_log$ and moves it to the end of $unacked_c$. The $\texttt{pull}(c)$ function models how the client $c$ learns about the next entry in the server log: it appends to $known_c$ the next operation in $server_log$ that is not yet part of $known_c$. If this operation is an echo of an operation previously executed by the same client $c$, we remove it from the $unacked_c$ log; the protocol ensures that in this case the operation is the first (oldest) one in $unacked_c$.

We model a client $c$ executing an operation $op$ with fences $fen \subseteq \{\mathsf{push}, \mathsf{pull}\}$ by $\texttt{exec}(c, op, fen)$. The body of $\texttt{exec}()$ is executed atomically, and only a single invocation of it can be in progress per client. At the beginning of $\texttt{exec}()$, we handle pull fences by repeatedly calling $\texttt{pull}(c)$ until the local $known_c$ matches $server_log$. At the end of $\texttt{exec}()$, we handle push fences by repeatedly calling $\texttt{push}(c)$ until all $pending_c$ operations have been processed by the server. At the core of $\texttt{exec}()$, we first compute the result of the operation by conjoining the logs $known_c$, $unacked_c$ and $pending_c$, stripping identifiers using stripIds and applying the sequential semantics of operations defined by eval (§2). We then append the operation to the $pending_c$ with a unique identifier generated by uniqueId. Since $op$ is evaluated on a log that includes $unacked_c$ and $pending_c$, the client is always guaranteed to observe its own operations, even before they are acknowledged by the server (the "read-your-writes" property [28]). Note that when $fen$ is empty, $\texttt{exec}(c, op, fen)$ returns immediately without communicating, so that in this case the protocol is partition-tolerant [14].

We only consider computations of the protocol that adhere to certain *fairness constraints*: every operation by a client eventually gets pushed to the server, every operation received by the server eventually gets pulled by any client and every invocation of $\texttt{exec}()$ terminates.

The set of histories $(E, \mathsf{so}, \mathsf{rt})$ allowed by GSC is defined by considering all possible computations of the above protocol. The invocations of $\texttt{exec}()$ define the set of events $E$, the order in which they are invoked on clients defines so, and two events are related by rt if the $\texttt{exec}()$ function of the former finishes before the $\texttt{exec}()$ function the latter starts. We denote the set of histories defined in this way HistGSC.

By systematically associating fences with operations in GSC we get various existing models as its special cases (Figure 1). If operations are executed without any fences, the GSC protocol exactly matches the one used to define GSP [10]. If every operation includes a pull fence, then the GSC protocol is isomorphic to one defining the *Total Store Order (TSO)* consistency model [22, 23]. In this case, operations are always evaluated based on an up-to-date state on the server, but are propagated to the server asynchronously. If every operation includes a push fence, then the GSC protocol is isomorphic to one defining a recently proposed *dual TSO* model [2]. In this case, all operations are pushed to the server immediately, but are evaluated on a client-local possibly stale state. If every operation includes both a pull and a push fence, then the GSC protocol produces exactly those histories that are linearizable [15] (we prove this in §C). Informally, in this case the total order in which the operations go into $server_log$ defines a linearization of the execution, which preserves the real-time order between the operations.

As a subcase of dual TSO, we also obtain a recently proposed *Ordered Sequential*

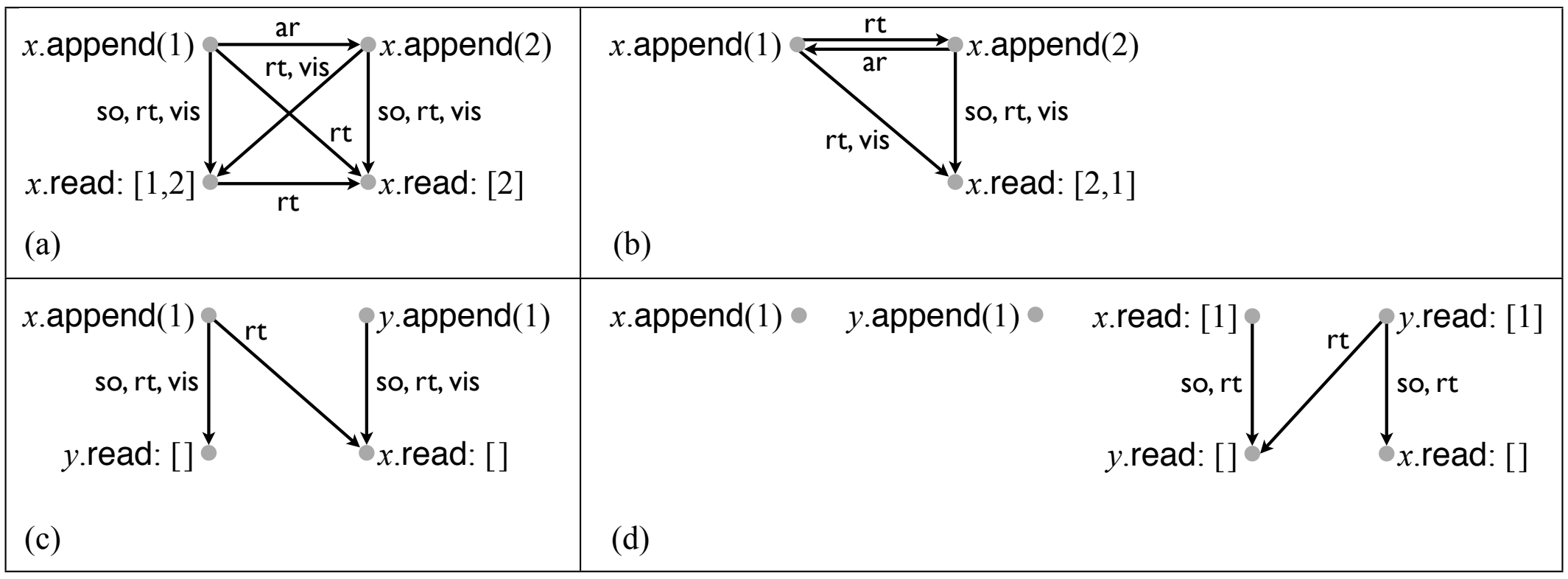

**Figure 3** Examples of histories and abstract executions. Events do not include fences unless explicitly noted. Events by the same client are related by the session order so and laid out vertically. Thus, there are two clients in (a-c) and four in (d).

*Consistency (OSC)* [21], which captures the semantics of coordination services such as ZooKeeper [17]. OSC assumes a partitioning of all operations into read-only and update operations: $\mathsf{Op} = \mathsf{OpReadOnly} \uplus \mathsf{OpUpdate}$. Read-only operations do not change the state of an object: for any operation $op$ and a sequence of operations $\xi$, we have $\mathsf{eval}(\xi, op) = \mathsf{eval}(\xi|_{\mathsf{OpUpdate}}, op)$, where $\xi|_{\mathsf{OpUpdate}}$ is the projection of $\xi$ onto OpUpdate. In our setting, OSC is defined by requiring that every operation include a push fence (like in dual TSO) and all updates additionally include a pull fence. Thus, update operations are evaluated on an up-to-date state, whereas read-only operations can be evaluated on a stale state. We prove the correspondence to the original OSC definition in §C.

With unrestricted fence placements, GSC is weaker than linearizability, as we illustrate by the example histories in Figures 3(a-c) (for now ignore the extra relations vis and ar). They use sequence objects $x$ and $y$ for which $\mathsf{eval}(\xi, \mathsf{read})$ returns the sequence of values in the append operations in $\xi$. The histories in Figures 3(a-c) can be produced by the GSC protocol, but are not linearizable: there does not exist a linearization of the events consistent with the real-time order and the sequential semantics of objects. In the following, we briefly describe how the GSC protocol produces these histories; the reader may wish to consult §A, where we describe the corresponding protocol computations in detail.

In history (a) the read by the second client does not see 1, even though it happens after the read by the first client that does see 1. In the GSC protocol this can happen if the second client does not pull append(1) from the server before executing the read. This history is disallowed if the read by the second client is executed with a pull fence: since the read by the first client returns $[1, 2]$, at the time the read is executed, 1 must be in *known* and, hence, on the server; then the pull fence ensures that the later read by the second client sees 1.

In history (b) the return value of the read is $[2, 1]$ even though append(1) finishes before append(2) starts. This can happen if the latter operation is pushed to the server before the former. This outcome is disallowed if append(1) is executed with a push fence, so that it is pushed to the server before the operation finishes.

In history (c) each read does not see the append by the other client; this is a variant of the *store buffering* anomaly, characteristic of TSO [23]. It can be produced by the GSC protocol if the appends are pushed to the server only after the reads execute. The history is disallowed if the appends include push fences and the reads pull fences.

Finally, history (d) is a variant of the *independent reads of independent writes* anomaly [7]

and cannot be produced by the GSC protocol. There two clients concurrently append 1 to different sequence objects $x$ and $y$. A third client sees the append to $x$, but not to $y$, and a fourth client sees the append to $y$, but not to $x$. Thus, from the perspective the latter two clients the updates to $x$ and $y$ happen in different orders. This outcome cannot happen in a GSC protocol computation, because there is a single order in which the append operations will be incorporated into the server log. If $x.\mathsf{append}(1)$ precedes $y.\mathsf{append}(1)$ in the log, then the read from $x$ in the fourth client cannot return $[\,]$; otherwise, the read from $y$ in the third client cannot return $[\,]$.

## 4 Equivalence between GSP, TSO and Dual TSO

We now establish a certain relationship between special cases of the GSC model: TSO [23] (all operations pull), dual TSO [2] (all operations push) and GSP [10] (operations neither pull nor push). We prove that the sets of histories allowed by these three models are the same modulo the real-time order, which means that the models are observationally equivalent to clients that cannot communicate out-of-band [3, 12].

Formally, for an event $e = (\iota, x, op, a, fen)$ let $\mathsf{mkPull}(e) = (\iota, x, op, a, \{\mathsf{pull}\})$ and $\mathsf{mkPush}(e) = (\iota, x, op, a, \{\mathsf{push}\})$. We lift mkPull and mkPush to sets of events and relations in the expected way. Let $\mathsf{EPush} = \{e \mid \mathsf{push} \in \mathsf{fences}(e)\}$ and $\mathsf{EPull} = \{e \mid \mathsf{pull} \in \mathsf{fences}(e)\}$.

▶ **Theorem 1.**

$$\begin{aligned}\forall E.\, \forall \mathsf{so}.\, E \cap (\mathsf{EPush} \cup \mathsf{EPull}) = \emptyset \implies\ & ((\exists \mathsf{rt}.\, (E, \mathsf{so}, \mathsf{rt}) \in \mathsf{HistGSC}) \iff \\ & (\exists \mathsf{rt}'.\, (\mathsf{mkPush}(E), \mathsf{mkPush}(\mathsf{so}), \mathsf{rt}') \in \mathsf{HistGSC}) \iff \\ & (\exists \mathsf{rt}''.\, (\mathsf{mkPull}(E), \mathsf{mkPull}(\mathsf{so}), \mathsf{rt}'') \in \mathsf{HistGSC})).\end{aligned}$$

We prove Theorem 1 in §7 and §C. According to it, any GSP computation of the protocol, where operations are propagated asynchronously both from clients to the server and from the server to clients, can be transformed into an equivalent-modulo-rt computation where operations can be propagated asynchronously in only one direction. While the equivalence between TSO and dual TSO has been established before [2], the result about GSP was only conjectured [10], and its proof is a contribution of the present paper. Like proofs of other results of ours, this one exploits the axiomatic specification of GSC that we present in §6.

If we take the real-time order into account and, hence, allow clients to communicate out-of-band, then GSP is strictly weaker than TSO and dual TSO, and the latter two are incomparable. In particular, the above theorem does not hold if we additionally require $\mathsf{rt}' = \mathsf{rt}$ or $\mathsf{rt}'' = \mathsf{rt}$. Indeed, as we noted in §3, the history in Figure 3(a) is allowed by GSP, but is disallowed if the operations pull; hence, it is disallowed by TSO. However, the history is allowed if all operations push and, hence, is allowed by dual TSO. The history in Figure 3(b) is similarly allowed by GSP, but is disallowed if all operations push; hence, it is disallowed by dual TSO. On the other hand, it is allowed if all operations pull and, hence, is allowed by TSO. Finally, even modulo real-time order, GSP, TSO and dual TSO are strictly weaker than linearizability [15]: the history in Figure 3(c) is allowed by these models, but is not linearizable no matter how we change the real-time order.

## 5 Composing GSC Objects

GSC is not a *composable* (aka *local*) property [15]: objects satisfying GSC may fail to provide this consistency guarantee when combined. Indeed, consider the history in Figure 3(d). It is easy to see that the projections of the history to events on objects $x$ or $y$ yield GSC histories:

e.g., the projection to $x$ can be produced by the GSC protocol if the rightmost client is slow to pull updates from the server. However, as we explained in §3, the overall history is not GSC. We now give a condition under which multiple objects each satisfying GSC behave such that the whole set of objects satisfies GSC. The condition requires using fences according to a certain discipline, formalised as follows. A history $\mathcal{H} = (E, \mathsf{so}, \mathsf{rt})$ is *well-fenced* if

$$\forall e, f \in E.\, e \xrightarrow{\mathsf{so}} f \land \mathsf{obj}(e) \neq \mathsf{obj}(f) \implies \exists e' \in \mathsf{EPush}.\, \exists f' \in \mathsf{EPull}.$$
$$\mathsf{obj}(e') = \mathsf{obj}(e) \land \mathsf{obj}(f') = \mathsf{obj}(f) \land e \xrightarrow{\mathsf{so}?} e' \xrightarrow{\mathsf{so}} f' \xrightarrow{\mathsf{so}?} f,$$

where $R?$ is the reflexive closure of $R$. The above condition requires that, when switching between different objects, a client pushes to the server the operations done on the old object and pulls from the server operations on the new object. Let us denote by $\mathcal{H}|_x$ the projection of $\mathcal{H}$ to events on an object $x$. The following theorem is our main result (proved in §8).

▶ **Theorem 2.** *For a well-fenced history $\mathcal{H}$, we have $(\forall x.\, \mathcal{H}|_x \in \mathsf{HistGSC}) \implies \mathcal{H} \in \mathsf{HistGSC}$.*

The theorem ensures that well-fenced clients interacting with multiple GSC services, implementing different objects, behave as though they are interacting with a single GSC service. Since our histories track the real-time order between events, this result holds even when clients can communicate out-of-band, without using GSC services. Programmers can thus ensure consistency when accessing multiple GSC services by placing fences according to the proposed discipline. Even though fences are expensive (in particular, not partition-tolerant), clients only incur this overhead when switching between different services. A client accessing the same service incurs no overhead.

For example, assume we make the upper reads in Figure 3(d) push and the lower reads pull. Then the projection of the history to $y$ is no longer GSC: since the lower read from $y$ happens after the upper read from $y$ and pulls operations from the server, it has to also observe 1. Hence, in this case the outcome shown in Figure 3(d) cannot happen when clients interact with multiple GSC services. (Actually, making the upper reads push is not required to ensure this, since they are read-only operations. Our results could be strengthened to incorporate such optimisations, but for simplicity we decided to treat all operations uniformly.)

As special cases of Theorem 2, we obtain novel criteria for composing TSO and dual TSO objects. Since in TSO all operations pull, we only need to require that a client pushes operations on an object before accessing a new one. Since in dual TSO all operations push, a client need only pull operations on the new object. As a subcase of dual TSO, we obtain the recently proposed criterion for composing OSC objects [21]. Recall that in OSC all operations push and update operations pull. Hence, in this case we require that a client start accessing a new object with an update operation. This can be ensured by adding dummy updates—a policy implemented by the ZooNet system [20] for composing ZooKeeper services [17]. Thus, our results generalise the compositionality criterion for OSC.

## 6 Axiomatic Specification

We now present the main technical tool we use to prove Theorems 1 and 2—an *axiomatic* specification of GSC, given in the style often used for consistency models in shared-memory [18] and distributed storage systems [8, 9]. It is based on the following notion. An *abstract execution* is a triple $\mathcal{A} = ((E, \mathsf{so}, \mathsf{rt}), \mathsf{vis}, \mathsf{ar})$, where: $(E, \mathsf{so}, \mathsf{rt})$ is a history; *visibility* $\mathsf{vis} \subseteq E \times E$ is a prefix-finite acyclic relation; and *arbitration* $\mathsf{ar} \subseteq E \times E$ is a prefix-finite total order such that $\mathsf{vis} \subseteq \mathsf{ar}$. Visibility and arbitration declaratively describe how the GSC protocol processes the operations in $E$. Given a computation of the protocol, we have $e \xrightarrow{\mathsf{vis}} f$ if, when a client executed the operation of $f$, the operation of $e$ was in one of its three local logs. We have $e \xrightarrow{\mathsf{ar}} f$ if the operation of $e$ preceded the one of $f$ in the server log. Figures 3(a-c) give examples of abstract executions (we omit some edges irrelevant for the following explanations).

To define the set of histories allowed by GSC, our specification constrains abstract executions using the *consistency axioms* in Figure 4, which declaratively describe guarantees the GSC protocol provides about operation processing and are explained in the following. In the axioms $R_1; R_2$ denotes the sequential composition of relations $R_1$ and $R_2$; we define $\mathsf{ctxt}_\mathcal{A}$ below. The axiomatic specification admits those histories that can be extended to an abstract execution satisfying the axioms. Denoting the latter set of executions $\mathsf{ExecGSC}$, the corresponding set of histories is

$$\mathsf{HistGSC}_{\mathrm{ax}} = \{\mathcal{H} \mid \exists \mathsf{vis}, \mathsf{ar}.\, (\mathcal{H}, \mathsf{vis}, \mathsf{ar}) \in \mathsf{ExecGSC}\}.$$

As the following shows, the axiomatic specification is equivalent to the operational one.

▶ **Theorem 3.** $\mathsf{HistGSC} = \mathsf{HistGSC}_{\mathrm{ax}}$.

RETVAL. $\forall e \in E.\, \mathsf{rval}(e) = \mathsf{eval}(\mathsf{ctxt}_\mathcal{A}(e), \mathsf{oper}(e))$.

RYW. $\mathsf{so} \subseteq \mathsf{vis}$.

MONOTONICVIEW. $\mathsf{vis} \,;\, \mathsf{so} \subseteq \mathsf{vis}$.

OBSERVEDVIS.

$\mathsf{ar}? \,;\, (\mathsf{vis} \setminus \mathsf{so}) \,;\, (\mathsf{rt} \cap (\mathsf{Event} \times \mathsf{EPull}))? \subseteq \mathsf{vis}$.

PUSHEDVIS. $\mathsf{ar}? \,;\, (\mathsf{rt}? \cap (\mathsf{EPush} \times \mathsf{EPull})) \subseteq \mathsf{vis}?$.

OBSERVEDAR. $(\mathsf{vis} \setminus \mathsf{so}) \,;\, \mathsf{rt} \subseteq \mathsf{ar}$.

PUSHEDAR. $\mathsf{rt} \cap (\mathsf{EPush} \times \mathsf{Event}) \subseteq \mathsf{ar}$.

EVENTUAL. $\forall e \in E.\, |\{f \in E \mid \neg(e \xrightarrow{\mathsf{vis}} f)\}| < \infty$.

**Figure 4** Axioms of the GSC model, constraining an execution $\mathcal{A} = ((E, \mathsf{so}, \mathsf{rt}), \mathsf{vis}, \mathsf{ar})$.

We now explain the axioms in Figure 4 and, on the way, give the key ideas for the proof of the "$\subseteq$" direction of the theorem, showing the *soundness* of the axiomatic specification. Consider a computation of the GSC protocol producing a history $\mathcal{H} = (E, \mathsf{so}, \mathsf{rt})$. To prove the soundness result, we extract $\mathsf{vis}$ and $\mathsf{ar}$ from the computation as described above and show that the resulting abstract execution satisfies all the axioms in Figure 4. RETVAL says that the result of an operation $e$ is computed by applying its sequential semantics to the sequence of operations given by $\mathsf{ctxt}_\mathcal{A}(e)$, which is obtained by arranging the operations invoked by the events in the set $\{f \mid f \xrightarrow{\mathsf{vis}} e \wedge \mathsf{obj}(e) = \mathsf{obj}(f)\}$ according to $\mathsf{ar}$. For example, the execution in Figure 3(b) satisfies RETVAL: the read returns $[2, 1]$ because both appends are visible to it and $x.\mathsf{append}(2) \xrightarrow{\mathsf{ar}} x.\mathsf{append}(1)$. RYW formalises the "read-your-writes" guarantee from §3: a client observes all operations it has executed before. MONOTONICVIEW similarly ensures that a client observes all operations it has observed before.

The axioms OBSERVEDVIS to PUSHEDAR are more subtle, and we thus give detailed justifications for their soundness. They constrain $\mathsf{vis}$ or $\mathsf{ar}$ based on the fact that, by a certain moment, a particular operation was guaranteed to have been pushed to the server. In OBSERVEDVIS and OBSERVEDAR this is the case because the operation was observed by a client other the one that that executed it (expressed in the axioms using $\mathsf{vis} \setminus \mathsf{so}$); in PUSHEDVIS and PUSHEDAR this is the case because the operation included a $\mathsf{push}$ fence (expressed using $\mathsf{EPush}$). In more detail, these axioms are justified as follows:

- OBSERVEDVIS. Assume $e_1 \xrightarrow{\mathsf{ar}?} e_2 \xrightarrow{\mathsf{vis}\setminus\mathsf{so}} e_3 \xrightarrow{\mathsf{rt}\cap(\mathsf{Event}\times\mathsf{EPull})?} e_4$. Since $e_2 \xrightarrow{\mathsf{vis}\setminus\mathsf{so}} e_3$, when a client executed $e_3$, it was aware of the event $e_2$ by a different client. The client could only find out about $e_2$ from the server, so by the time $e_3$ finished, $e_2$ was on the server.

Since $e_1 \xrightarrow{\mathsf{ar}?} e_2$, so was $e_1$. If $e_3 = e_4$, then the client executing this event was also aware of $e_1$, since clients pull operations in the order of the server log. Hence, $e_1 \xrightarrow{\mathsf{vis}} e_4$. If $e_3 \xrightarrow{\mathsf{rt} \cap (\mathsf{Event} \times \mathsf{EPull})} e_4$, then after $e_3$ finished, the client executing $e_4$ pulled all updates from the server, which must have included $e_1$. Hence, $e_1 \xrightarrow{\mathsf{vis}} e_4$ again.

- PushedVis. Assume $e_1 \xrightarrow{\mathsf{ar}?} e_2 \xrightarrow{\mathsf{rt}?} e_3$, $e_2 \in \mathsf{EPush}$ and $e_3 \in \mathsf{EPull}$. Since $e_2 \in \mathsf{EPush}$, $e_2$ was on the server after its operation finished. Since $e_1 \xrightarrow{\mathsf{ar}?} e_2$, so was $e_1$. If $e_1 = e_3$, we trivially have $e_1 \xrightarrow{\mathsf{vis}?} e_3$. Otherwise, since $e_2 \xrightarrow{\mathsf{rt}?} e_3$, $e_1$ was also on the server before $e_3$ started. Since $e_3 \in \mathsf{EPull}$, $e_3$ pulled all operations from the server, including $e_1$. Hence, $e_1 \xrightarrow{\mathsf{vis}} e_3$.
- ObservedAr. Assume $e_1 \xrightarrow{\mathsf{vis} \setminus \mathsf{so}} e_2 \xrightarrow{\mathsf{rt}} e_3$. Since $e_1 \xrightarrow{\mathsf{vis} \setminus \mathsf{so}} e_2$, $e_1$ must have been on the server by the time $e_2$ finished. Since $e_2 \xrightarrow{\mathsf{rt}} e_3$, $e_3$ started after $e_2$ finished and thus must follow $e_1$ in the server log. Hence, $e_1 \xrightarrow{\mathsf{ar}} e_3$.
- PushedAr. Assume $e_1 \xrightarrow{\mathsf{rt}} e_2$ and $e_1 \in \mathsf{EPush}$. Then $e_1$ was pushed to the server before $e_2$ started. Hence, $e_2$ was pushed onto the server after $e_1$, so that $e_1 \xrightarrow{\mathsf{ar}} e_2$.

Finally, the Eventual axiom guarantees that an event $e$ can be invisible to at most finitely many other events $f$. Its soundness is ensured by the fairness constraints in the GSC protocol (§3). The axioms imply more properties of the relations in an execution.

▶ **Proposition 4.** *If $\mathcal{A}$ satisfies* MonotonicView *and* ObservedVis*, then $\mathsf{vis}_{\mathcal{A}}$ is transitive. If $\mathcal{A}$ satisfies* ObservedAr*, then $\mathsf{vis}_{\mathcal{A}} \cup \mathsf{rt}_{\mathcal{A}}$ is acyclic.*

The executions in Figures 3(a-c) satisfy all the axioms. On the other hand, the history in Figure 3(d) cannot be extended to an execution satisfying the axioms. Indeed, for the return values of the upper reads to be consistent with RetVal, we must have $x.\mathsf{append}(1) \xrightarrow{\mathsf{vis}} x.\mathsf{read} : [1]$ and $y.\mathsf{append}(1) \xrightarrow{\mathsf{vis}} y.\mathsf{read} : [1]$. Arbitration has to order the two appends one way or another. If, for example, we have $x.\mathsf{append}(1) \xrightarrow{\mathsf{ar}} y.\mathsf{append}(2)$, then by ObservedVis we must also have $x.\mathsf{append}(1) \xrightarrow{\mathsf{vis}} x.\mathsf{read} : [\,]$, contradicting RetVal.

Recall from §3 that GSC disallows the history in Figure 3(a) if the read in the second client is a pull. Accordingly, there is no abstract execution that extends the resulting history and satisfies the axioms: by ObservedVis, in such an execution we would have $x.\mathsf{append}(1) \xrightarrow{\mathsf{vis}} x.\mathsf{read} : [2]$, contradicting RetVal. Similarly, there is no execution that extends the history in Figure 3(b) assuming $x.\mathsf{append}(1)$ is a push. This is because by PushedAr in such an execution we must have $x.\mathsf{append}(1) \xrightarrow{\mathsf{ar}} x.\mathsf{append}(2)$, so that by RetVal the read must return $[1, 2]$. Finally, there is no execution for the history in Figure 3(c) assuming the appends push and the reads pull: by PushedVis we must have $x.\mathsf{append}(1) \xrightarrow{\mathsf{vis}} x.\mathsf{read} : [\,]$, contradicting RetVal.

As follows from the "$\supseteq$" direction of Theorem 3, the axioms in Figure 4 are also *complete*: given an abstract execution $(\mathcal{H}, \mathsf{vis}, \mathsf{ar})$, we can construct a computation of the GSC protocol producing the history $\mathcal{H}$. Due to space constraints, we defer the detailed proof of Theorem 3 to §B. The completeness part of the proof is nontrivial, but uses similar techniques to the proof of the compositionality criterion that we present in §8.

## 7 Proof of Model Equivalence

As a simple illustration of the use of the axiomatic specification of GSC, we prove the first "$\Longleftrightarrow$" in Theorem 1, showing that GSP and dual TSO are equivalent modulo real-time order (the rest of the proof is given in §C). Consider $E$ and $\mathsf{so}$ such that $E \cap (\mathsf{EPush} \cup \mathsf{EPull}) = \emptyset$.

*The "$\Longleftarrow$" direction.* It is easy to see that

$$\forall \mathsf{rt}.\, (\mathsf{mkPush}(E), \mathsf{mkPush}(\mathsf{so}), \mathsf{mkPush}(\mathsf{rt})) \in \mathsf{HistGSC} \implies (E, \mathsf{so}, \mathsf{rt}) \in \mathsf{HistGSC},$$

since erasing fences from events does not invalidate any axioms.

*The "$\Longrightarrow$" direction.* Assume $\mathsf{rt}$ such that $(E, \mathsf{so}, \mathsf{rt}) \in \mathsf{HistGSC}$. Then for some $\mathsf{vis}$ and $\mathsf{ar}$ we have $\mathcal{A} \triangleq ((E, \mathsf{so}, \mathsf{rt}), \mathsf{vis}, \mathsf{ar}) \in \mathsf{ExecGSC}$. Let $\mathsf{rt}' = \mathsf{mkPush}(\mathsf{ar})$. Then

$$\mathcal{A}' \triangleq ((\mathsf{mkPush}(E), \mathsf{mkPush}(\mathsf{so}), \mathsf{rt}'), \mathsf{mkPush}(\mathsf{vis}), \mathsf{mkPush}(\mathsf{ar}))$$

is an abstract execution. Further, since $\mathcal{A}$ satisfies all GSC axioms, so does $\mathcal{A}'$. In particular, $\mathcal{A}'$ satisfies ObservedVis and PushedVis because $\mathsf{mkPush}(E) \cap \mathsf{EPull} = \emptyset$, and ObservedAr and PushedAr by the choice of $\mathsf{rt}'$. This completes the proof.

Thus, our axiomatic specification allows easily proving the above model equivalence by picking a witness for the real-time order and checking axiom validity. Such a proof would be much more challenging with the operational specification, as it would require devising a nontrivial transformation of one execution of the GSC protocol into another.

## 8 Proof of the Compositionality Criterion

We next show how to use our axiomatic specification of the GSC model to prove Theorem 2. Here we give only the key ideas and defer the complete proof to §D. Consider a well-fenced history $\mathcal{H} = (E, \mathsf{so}, \mathsf{rt})$ such that $\forall x.\, \mathcal{H}|_x \in \mathsf{HistGSC}$. Then for any $x$ there is an execution $\mathcal{A}_x = (\mathcal{H}|_x, \mathsf{vis}_x, \mathsf{ar}_x) \in \mathsf{ExecGSC}$. We need to show $\mathcal{H} \in \mathsf{HistGSC}$, to which end we construct an execution $\mathcal{A} = (\mathcal{H}, \mathsf{vis}, \mathsf{ar}) \in \mathsf{ExecGSC}$.

Let $\mathsf{so}_0 = \bigcup_{x \in \mathsf{Obj}} \mathsf{so}_{\mathcal{H}|_x}$, $\mathsf{vis}_0 = \bigcup_{x \in \mathsf{Obj}} \mathsf{vis}_x$ and $\mathsf{ar}_0 = \bigcup_{x \in \mathsf{Obj}} \mathsf{ar}_x$. It is reasonable to expect $\mathsf{vis}$ and $\mathsf{ar}$ to extend the corresponding per-object orders in $\mathcal{A}_x$, so we should have $\mathsf{vis}_0 \subseteq \mathsf{vis}$ and $\mathsf{ar}_0 \subseteq \mathsf{ar}$. The most difficult part is to construct $\mathsf{ar}$; once this is done, we construct $\mathsf{vis}$ as the smallest relation containing $\mathsf{vis}_0$ that is a solution to the system of inequalities given by the axioms RYW-PushedVis in Figure 4. The following lemma gives a closed form for this solution. Let $\mathsf{Id} = \{(e, e) \mid e \in E\}$.

▶ **Lemma 5.** *Given any arbitration order* $\mathsf{ar} \supseteq \mathsf{ar}_0$*, the relation*

$$\mathsf{vis} = \mathsf{so} \cup (\mathsf{ar}?\,;(\mathsf{vis}_0 \setminus \mathsf{so})\,;(\mathsf{rt} \cap (\mathsf{Event} \times \mathsf{EPull}))?\,;\mathsf{so}?) \cup ((\mathsf{ar}?\,;(\mathsf{rt}? \cap (\mathsf{EPush} \times \mathsf{EPull}))\,;\mathsf{so}?) \setminus \mathsf{Id})$$

*is the smallest one such that* $\mathsf{vis}_0 \subseteq \mathsf{vis}$ *and* $(\mathcal{H}, \mathsf{vis}, \mathsf{ar})$ *satisfies* RYW-PushedVis.

The first component of $\mathsf{vis}$ is meant to validate RYW, the second ObservedVis and the third PushedVis. Appending $\mathsf{so}?$ at the end of the last two components validates MonotonicView.

We now describe the construction of $\mathsf{ar}$. This order needs to include several relations. Since $\mathsf{vis}_0 \subseteq \mathsf{vis}$ and $\mathcal{A}$ should satisfy ObservedAr, we must have $(\mathsf{vis}_0 \setminus \mathsf{so})\,;\mathsf{rt} \subseteq \mathsf{ar}$. Since $\mathcal{A}$ should satisfy PushedAr we must have $\overline{\mathsf{rt}} \triangleq \mathsf{rt} \cap (\mathsf{EPush} \times \mathsf{Event}) \subseteq \mathsf{ar}$. Since $\mathcal{A}$ should satisfy RYW and $\mathsf{vis} \subseteq \mathsf{ar}$, we must have $\mathsf{so} \subseteq \mathsf{ar}_0$. Finally, for $\mathcal{A}$ to satisfy RetVal, $\mathsf{ar}$ should include one more relation that is more subtle. We illustrate the need for it using the example in Figure 5. Assume that we have the solid edges in the figure. If we arbitrate between the two appends as shown by the dashed edge $f \xrightarrow{\mathsf{ar}} e$, then according to the construction in Lemma 5 we will also have the dashed edge $f \xrightarrow{\mathsf{vis}} g$ (needed for $\mathcal{A}$ to satisfy ObservedVis). But then the resulting $\mathcal{A}$ will violate RetVal. We therefore include the following relation into $\mathsf{ar}$, which ensures that such situations do not happen:

$$e \prec f \iff \exists g.\, \mathsf{obj}(f) = \mathsf{obj}(g) \wedge (f, g) \notin \mathsf{vis}_0 \wedge$$
$$(e, g) \in (\mathsf{vis}_0 \setminus \mathsf{so}) ; (\mathsf{rt} \cap (\mathsf{Event} \times \mathsf{EPull})) ; \mathsf{so}_0? \cup (\mathsf{rt} \cap (\mathsf{EPush} \times \mathsf{EPull})) ; \mathsf{so}_0?.$$

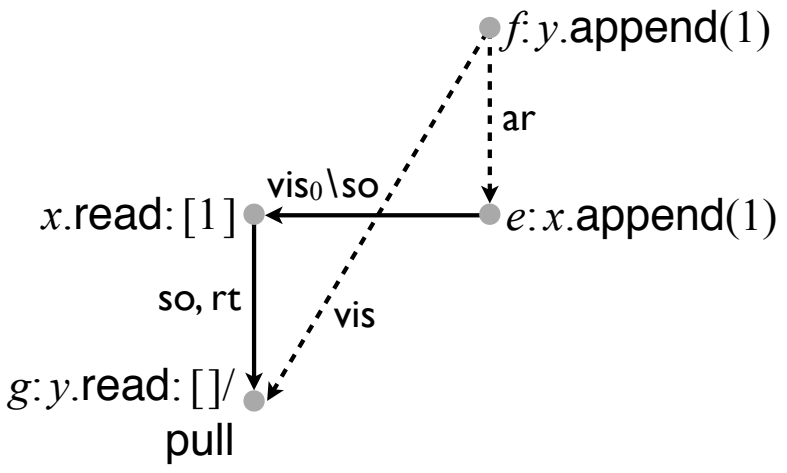

**Figure 5** Motivation for $\prec$.

If $e \prec f$, then adding an edge $f \xrightarrow{\mathsf{ar}} e$ would create a visibility edge $f \xrightarrow{\mathsf{vis}} g$ between events on the same object that is not in $\mathsf{vis}_0$. Note that the expression covering $(e, g)$ above is more specific than the one in Lemma 5: we have $\mathsf{so}_0$ instead of $\mathsf{so}$, and $\mathsf{rt}$ must be used. This is crucial for the proof (specifically, Lemma 6 below) and, as we show, is still sufficient to validate RetVal because the history $\mathcal{H}$ is well-fenced.

Thus, we need to construct an $\mathsf{ar}$ that includes $R \triangleq \overline{\mathsf{rt}} \cup \mathsf{so} \cup \mathsf{ar}_0 \cup ((\mathsf{vis}_0 \setminus \mathsf{so}) ; \mathsf{rt}) \cup \prec$. For this to be possible, $R$ has to be acyclic.

▶ **Lemma 6.** $\overline{\mathsf{rt}} \cup \mathsf{so} \cup \mathsf{ar}_0 \cup ((\mathsf{vis}_0 \setminus \mathsf{so}) ; \mathsf{rt}) \cup \prec$ *is acyclic.*

Establishing this lemma is the most subtle part of the proof. To do this, we construct a closed-form expression covering the transitive closure of $R$.

▶ **Lemma 7.**

$$(\overline{\mathsf{rt}} \cup \mathsf{so} \cup \mathsf{ar}_0 \cup ((\mathsf{vis}_0 \setminus \mathsf{so}) ; \mathsf{rt}) \cup \prec)^+$$
$$= (\overline{\mathsf{rt}} \cup \mathsf{so} \cup \mathsf{ar}_0 \cup ((\mathsf{vis}_0 \setminus \mathsf{so}) ; \mathsf{rt}))^+ \cup (\prec \cup\, \mathsf{ar}_0 ; \prec) ; (\overline{\mathsf{rt}} \cup \mathsf{so} \cup \mathsf{ar}_0 \cup ((\mathsf{vis}_0 \setminus \mathsf{so}) ; \mathsf{rt}))^* \quad \textit{and}$$

$$(\overline{\mathsf{rt}} \cup \mathsf{so} \cup \mathsf{ar}_0 \cup ((\mathsf{vis}_0 \setminus \mathsf{so}) ; \mathsf{rt}))^+$$
$$\subseteq \overline{\mathsf{rt}} \cup \mathsf{ar}_0 \cup (\mathsf{ar}_0 ; \overline{\mathsf{rt}}) \cup (\overline{\mathsf{rt}} ; \mathsf{ar}_0) \cup (\mathsf{ar}_0 ; \overline{\mathsf{rt}} ; \mathsf{ar}_0) \cup ((\mathsf{vis}_0 \setminus \mathsf{so}) ; \mathsf{rt}) \cup$$
$$(\mathsf{ar}_0 ; ((\mathsf{vis}_0 \setminus \mathsf{so}) ; \mathsf{rt})) \cup (((\mathsf{vis}_0 \setminus \mathsf{so}) ; \mathsf{rt}) ; \mathsf{ar}_0) \cup (\mathsf{ar}_0 ; ((\mathsf{vis}_0 \setminus \mathsf{so}) ; \mathsf{rt}) ; \mathsf{ar}_0).$$

The proof Lemma 7 relies on establishing that components of $R$ satisfy various algebraic properties, some of which exploit the fact that the history $\mathcal{H}$ is well-fenced. For example, we prove that $\prec$ is a strict partial order, i.e., transitive and irreflexive.

To prove Lemma 6, it is thus sufficient to prove that the relation covering $R^+$ in Lemma 7 is irreflexive. This relation describes only particular paths in $R$ of length at most 5. Its irreflexivity is then established by a case analysis on these paths.

Using Lemma 6, we can extend $R$ to a prefix-finite total order, which we take as $\mathsf{ar}$; then $\mathsf{vis}$ is defined by Lemma 5. We can then show that $\mathsf{vis}$ defined in this way is prefix-finite, acyclic and $\mathsf{vis} \subseteq \mathsf{ar}$, so that $\mathcal{A} = (\mathcal{H}, \mathsf{vis}, \mathsf{ar})$ is an abstract execution. By Lemma 5, $\mathcal{A}$ satisfies RYW-PushedVis. It satisfies PushedAr because $\overline{\mathsf{rt}} \subseteq \mathsf{ar}$, and it is also easy to check that it satisfies ObservedAr.

We next argue that $\mathcal{A}$ satisfies RetVal, which exploits the particular way in which we constructed $\mathsf{ar}$. To this end, we show that for any object $x$ we have $\mathsf{vis}|_x = \mathsf{vis}_x$, where $\mathsf{vis}|_x$ is the projection of $\mathsf{vis}$ to events on $x$. Then since for any $x$ we have $\mathsf{ar}_x \subseteq \mathsf{ar}$ and $\mathcal{A}_x$ satisfies RetVal, so does $\mathcal{A}$. Since $\mathsf{vis}_x \subseteq \mathsf{vis}$ by construction, we only need to show $\mathsf{vis}|_x \subseteq \mathsf{vis}_x$. Consider arbitrary $f, g \in E$ such that $\mathsf{obj}(f) = \mathsf{obj}(g) = x$ and $f \xrightarrow{\mathsf{vis}} g$. To show $f \xrightarrow{\mathsf{vis}_x} g$ our proof considers several cases corresponding to which of the components of the union defining $\mathsf{vis}$ in Lemma 5 the edge $(f, g)$ belongs to. For illustration, here we only consider a single case when $(f, g)$ comes from the following instance of the second component of the union, which uses an $\mathsf{rt}$ edge: $(f, g) \in \mathsf{ar}? ; (\mathsf{vis}_0 \setminus \mathsf{so}) ; (\mathsf{rt} \cap (\mathsf{Event} \times \mathsf{EPull})) ; \mathsf{so}?$. Then for some $g'$ we have

$$f \xrightarrow{\mathsf{ar}?;(\mathsf{vis}_0 \setminus \mathsf{so});(\mathsf{rt} \cap (\mathsf{Event} \times \mathsf{EPull}))} g' \xrightarrow{\mathsf{so}?} g.$$

Figure 5 illustrates the case when $g' = g$. If $\mathsf{obj}(g') \neq \mathsf{obj}(g)$, then since the history $\mathcal{H}$ is well-fenced, for some $g'' \in \mathsf{EPull}$ we have $g' \xrightarrow{\mathsf{so}} g'' \xrightarrow{\mathsf{so}_0?} g$. Since $\mathsf{so} \subseteq \mathsf{rt}$, this implies $g' \xrightarrow{\mathsf{rt} \cap (\mathsf{Event} \times \mathsf{EPull})} g'' \xrightarrow{\mathsf{so}_0?} g$. Hence,

$$f \xrightarrow{\mathsf{ar}?;(\mathsf{vis}_0 \setminus \mathsf{so});(\mathsf{rt} \cap (\mathsf{Event} \times \mathsf{EPull}))} g'' \xrightarrow{\mathsf{so}_0?} g. \tag{1}$$

If $\mathsf{obj}(g') = \mathsf{obj}(g)$, then $g' \xrightarrow{\mathsf{so}_0?} g$ and we again have (1) for $g'' = g'$. Thus, in all cases (1) holds for some $g''$. Then for some $e$ we have

$$f \xrightarrow{\mathsf{ar}?} e \xrightarrow{(\mathsf{vis}_0 \setminus \mathsf{so});(\mathsf{rt} \cap (\mathsf{Event} \times \mathsf{EPull}))} g'' \xrightarrow{\mathsf{so}_0?} g.$$

Now if $\neg(f \xrightarrow{\mathsf{vis}_x} g)$, then $e \prec f$, contradicting $\prec \subseteq \mathsf{ar}$. Hence, $f \xrightarrow{\mathsf{vis}_x} g$, as required.

Thus, $\mathcal{A}$ satisfies all GSC axioms except for possibly EVENTUAL. Since $\forall x.\, \mathsf{vis}|_x = \mathsf{vis}_x$ and $\mathcal{A}_x$ satisfies EVENTUAL, we have

$$\forall e \in E.\, |\{f \in E \mid \mathsf{obj}(e) = \mathsf{obj}(f) \wedge \neg(e \xrightarrow{\mathsf{vis}} f)\}| < \infty, \tag{2}$$

i.e., an event $e$ cannot be invisible to infinitely many events $f$ on the same object. Then, as the following lemma shows, we can extend vis so as to validate EVENTUAL without invalidating any of the other axioms.

▶ **Lemma 8.** *Let $\mathcal{H} = (E, \mathsf{so}, \mathsf{rt})$ and $\mathcal{A} = (\mathcal{H}, \mathsf{vis}, \mathsf{ar})$ be an execution that satisfies all GSC axioms except for possibly* EVENTUAL*. Assume (2) holds. Then there exists $\mathsf{vis}' \supseteq \mathsf{vis}$ such that $(\mathcal{H}, \mathsf{vis}', \mathsf{ar}) \in \mathsf{ExecGSC}$.*

We thus construct an execution $(\mathcal{H}, \mathsf{vis}', \mathsf{ar}) \in \mathsf{ExecGSC}$, which shows that $\mathcal{H} \in \mathsf{HistGSC}$ and thereby establishes Theorem 2.

The axiomatic specification of GSC plays an important role in the above proof. It allows us to concisely state constraints that the global order on operations represented by ar needs to satisfy for the global execution to be GSC. We can then show that the desired global order exists by proving algebraic properties over relations, as exemplified by Lemma 7.

## 9 Related Work and Discussion

Lev-Ari et al. [21] have proposed a criterion for composing objects providing Ordered Sequential Consistency (OSC), which is a special case of our results (§5). In comparison to them, we handle a more complex consistency model, which requires a different proof approach: specifying the consistency model axiomatically and reasoning about it using algebraic techniques. Lev-Ari et al. have also implemented their criterion in a library for composing ZooKeeper instances and showed that it has a competitive performance [20]. We hope that our results will enable similar practical implementations for systems providing other consistency models from the family we considered. In particular, the implementation of GSP in Orleans [5] provides only per-object consistency guarantees, and our results should allow its clients to use multiple objects while preserving the consistency model.

There are other widely used consistency models that are in general non-composable, such as sequential consistency [19]. Perrin et al. [24] proposed conditions on the use of sequentially consistent concurrent objects under which a composition of multiple objects stays sequentially consistent. Our compositionality result is similar in spirit, but handles a family of more complex consistency models implemented in modern systems [10, 17, 22]. Vitenberg and Friedman [29] showed that combining sequential consistency with any composable property yields a non-composable property. Our compositionality criterion does not contradict this result, since well-fencedness of histories is not a composable property.

Our operational specification of the GSC model generalizes the GSP protocol [10], with significant differences. First, GSP allows only pure read and update operations, while GSC permits mixed operations that both modify the state and return a value to the caller. Second, GSP does not support push and pull fences that are attached to operations. Rather, its original proposal [10] investigated stronger synchronization primitives, such as standalone fences and transactions, which cannot be used to define TSO, dual TSO and OSC as special cases. Therefore, GSP is unsuitable to serve as a unifying model that clarifies the relationship between these instances.

Axiomatic specifications have been previously proposed for consistency models in shared-memory [18, 22] and distributed storage systems [8, 9]. Our GSC specification uses the same framework as for the latter. Researchers have proposed axiomatic specifications for TSO-like models and proved their equivalence to operational ones [16, 22]. However, our specifications are the first to formalise the role of the real-time order in distinguishing between these models. Including real-time order into axiomatic models [8] is important in a distributed setting because of the possibility of out-of-band communication between clients; without this one cannot safely substitute implementations for specifications [3, 12].

We have exploited the axiomatic specification of GSC to establish a compositionality criterion and an equivalence between GSP and TSO/dual TSO. However, axiomatic specifications of consistency models have been shown useful to obtain other kinds of results, such as criteria for *robustness*—checking when an application running on a weak consistency model behaves as if it runs on a strong one [4, 26]. We hence hope that our specifications will allow obtaining such results for consistency models with global operation sequencing.

**Acknowledgements.** We thank Idit Keidar, Kfir Lev-Ari and Matthieu Perrin for helpful comments. Gotsman was supported by an ERC Starting Grant RACCOON.

## APPENDIX

# A Execution Examples

For illustration, we now give several detailed execution traces of the GSC protocol (Figure 2) that match the histories and abstract executions in Figure 3. Our notation has time going from top to bottom. There are two clients, $A$ and $B$, whose state and transitions are aligned to the left and right, while server state and communication appears in the middle. All state annotations are enclosed in brackets.

## A.1 Example (a)

| Client $A$ | Server | Client $B$ |
| --- | --- | --- |
| $\begin{bmatrix} known_A = [\,] \\ unacked_A = [\,] \\ pending_A = [\,] \end{bmatrix}$ | $[server_log = [\,]]$ | $\begin{bmatrix} known_B = [\,] \\ unacked_B = [\,] \\ pending_B = [\,] \end{bmatrix}$ |
| **start** $\texttt{exec}(A, x.\mathsf{append}(1), \emptyset)$ | | **start** $\texttt{exec}(B, x.\mathsf{append}(2), \emptyset)$ |
| $[pending_A = x.\mathsf{append}(1)]$ | | $[pending_B = x.\mathsf{append}(2)]$ |
| **return** | | **return** |
| | $\texttt{push}(A)$ | |
| $\begin{bmatrix} unacked_A = x.\mathsf{append}(1) \\ pending_A = [\,] \end{bmatrix}$ | $[server_log = x.\mathsf{append}(1)]$ | |
| | $\texttt{push}(B)$ | |
| | $\begin{bmatrix} server_log = \begin{matrix} x.\mathsf{append}(1) \\ x.\mathsf{append}(2) \end{matrix} \end{bmatrix}$ | $\begin{bmatrix} unacked_B = x.\mathsf{append}(2) \\ pending_B = [\,] \end{bmatrix}$ |
| | $\texttt{pull}(A)$ | |
| $\begin{bmatrix} known_A = x.\mathsf{append}(1) \\ unacked_A = [\,] \end{bmatrix}$ | | |
| | $\texttt{pull}(A)$ | |
| $\begin{bmatrix} known_A = \begin{matrix} x.\mathsf{append}(1) \\ x.\mathsf{append}(2) \end{matrix} \end{bmatrix}$ | | |
| **start** $\texttt{exec}(A, x.\mathsf{read}(), \emptyset)$ | | |
| **return** $[1, 2]$ | | |
| | | **start** $\texttt{exec}(B, x.\mathsf{read}(), \emptyset)$ |
| | | **return** $[2]$ |
| | $\texttt{pull}(B)$ | |
| | $\texttt{pull}(B)$ | |
| | | $\begin{bmatrix} known_B = \begin{matrix} x.\mathsf{append}(1) \\ x.\mathsf{append}(2) \end{matrix} \\ unacked_B = [\,] \end{bmatrix}$ |

## A.2 Example (b)

$$\begin{bmatrix} known_A = [\,] \\ unacked_A = [\,] \\ pending_A = [\,] \end{bmatrix} \qquad [server_log = [\,]] \qquad \begin{bmatrix} known_B = [\,] \\ unacked_B = [\,] \\ pending_B = [\,] \end{bmatrix}$$

**start** $\texttt{exec}(A, x.\mathsf{append}(1), \emptyset)$

**return**

**start** $\texttt{exec}(B, x.\mathsf{append}(2), \emptyset)$

**return**

$$\begin{bmatrix} known_A = [\,] \\ unacked_A = [\,] \\ pending_A = x.\mathsf{append}(1) \end{bmatrix} \qquad [server_log = [\,]] \qquad \begin{bmatrix} known_B = [\,] \\ unacked_B = [\,] \\ pending_B = x.\mathsf{append}(2) \end{bmatrix}$$

$\texttt{push}(B)$
$\texttt{push}(A)$

$$\begin{bmatrix} unacked_A = x.\mathsf{append}(1) \\ pending_A = [\,] \end{bmatrix} \qquad \begin{bmatrix} server_log = \begin{array}{c} x.\mathsf{append}(2) \\ x.\mathsf{append}(1) \end{array} \end{bmatrix} \qquad \begin{bmatrix} unacked_B = x.\mathsf{append}(2) \\ pending_B = [\,] \end{bmatrix}$$

$\texttt{pull}(B)$
$\texttt{pull}(B)$
$\texttt{pull}(A)$
$\texttt{pull}(A)$

$$\begin{bmatrix} known_A = \begin{array}{c} x.\mathsf{append}(2) \\ x.\mathsf{append}(1) \end{array} \\ unacked_A = [\,] \\ pending_A = [\,] \end{bmatrix} \qquad \begin{bmatrix} known_B = \begin{array}{c} x.\mathsf{append}(2) \\ x.\mathsf{append}(1) \end{array} \\ unacked_B = [\,] \\ pending_B = [\,] \end{bmatrix}$$

**start** $\texttt{exec}(B, x.\mathsf{read}(), \emptyset)$

**return** $[2, 1]$

## A.3 Example (c)

$$\begin{bmatrix} known_A = [\,] \\ unacked_A = [\,] \\ pending_A = [\,] \end{bmatrix} \qquad [server_log = [\,]] \qquad \begin{bmatrix} known_B = [\,] \\ unacked_B = [\,] \\ pending_B = [\,] \end{bmatrix}$$

**start** $\texttt{exec}(A, x.\mathsf{append}(1), \emptyset)$ — **start** $\texttt{exec}(B, y.\mathsf{append}(1), \emptyset)$

$[pending_A = x.\mathsf{append}(1)]$ — $[pending_B = y.\mathsf{append}(1)]$

**return**

**start** $\texttt{exec}(A, y.\mathsf{read}(), \emptyset)$

**return** (B)

**return** $[\,]$ (A)

**start** $\texttt{exec}(B, x.\mathsf{read}(), \emptyset)$

**return** $[\,]$

$\texttt{push}(A)$
$\texttt{push}(B)$
$\texttt{pull}(A)$
$\texttt{pull}(A)$
$\texttt{pull}(B)$
$\texttt{pull}(B)$

$$\begin{bmatrix} known_A = \begin{matrix} x.\mathsf{append}(1) \\ y.\mathsf{append}(1) \end{matrix} \\ unacked_A = [\,] \\ pending_A = [\,] \end{bmatrix} \quad \begin{bmatrix} server_log = \begin{matrix} x.\mathsf{append}(1) \\ y.\mathsf{append}(1) \end{matrix} \end{bmatrix} \quad \begin{bmatrix} known_B = \begin{matrix} x.\mathsf{append}(1) \\ y.\mathsf{append}(1) \end{matrix} \\ unacked_B = [\,] \\ pending_B = [\,] \end{bmatrix}$$

## B Proof of Theorem 3

We first prove that $\mathsf{HistGSC} \subseteq \mathsf{HistGSC}_{\mathrm{ax}}$. Take $\mathcal{H} = (E, \mathsf{so}, \mathsf{rt}) \in \mathsf{HistGSC}$. Then there exists an computation of the GSC protocol producing this history. We extract $\mathsf{vis}$ and $\mathsf{ar}$ from the computation as described in §6 and show that the resulting abstract execution $\mathcal{A} = (\mathcal{H}, \mathsf{vis}, \mathsf{ar})$ satisfies all the axioms in Figure 4. The arguments for the validity of ObservedVis, PushedVis, ObservedAr and PushedAr were given in §6, so here we give the arguments for the remaining axioms.

- RYW. Assume $e_1 \xrightarrow{\mathsf{so}} e_2$. Then $e_1$ and $e_2$ are executed by the same client in this order. Hence, $e_1$ is in one of the logs of the client when $e_2$ is executed, so $e_1 \xrightarrow{\mathsf{vis}} e_2$.
- MonotonicView. $e_1 \xrightarrow{\mathsf{vis}} e_2 \xrightarrow{\mathsf{so}} e_3$. Then $e_2$ and $e_3$ are executed by the same client in this order. At the time when $e_2$ is executed, $e_1$ is in one of the logs of this client. Hence, it is also in one of its logs when $e_3$ is executed.
- Eventual. This follows from the fairness constraints on protocol computations that we required in §3.
- RetVal. This follows from the way *result* is computed in `exec`() and from the fact that the order of entries in $known_c \cdot unacked_c \cdot pending_c$ in any client $c$ is consistent with their order in *server_log*, which corresponds to $\mathsf{ar}$.

We now prove $\mathsf{HistGSC}_{\mathrm{ax}} \subseteq \mathsf{HistGSC}$. Consider an execution $\mathcal{A} = (\mathcal{H}, \mathsf{vis}, \mathsf{ar}) \in \mathsf{ExecGSC}$, where $\mathcal{H} = (E, \mathsf{so}, \mathsf{rt})$. We need to construct a computation $\mathcal{C}$ of the GSC protocol producing the history $\mathcal{H}$.

In this computation the bodies of `exec`(), `push`() and `pull`() are executed atomically, though calls to them and returns from them are separate transitions, executed as we define below. The points at which we invoke `push`() and `pull`() are determined by invocations of `exec`() so that pushing and pulling of operations is delayed for as long as possible. Namely, we follow the following rules:

1. Right before a client calls `exec`(), generating an event $e \in E$, we execute enough `push`() functions at other clients and `pull`() functions at the current client to pull into the current client's logs all events $f \in E$ such that $f \xrightarrow{\mathsf{vis} \setminus \mathsf{so}} e$. Note that this may also involve pulling events by the current client from the server.
2. Whenever we execute `push`() at a client to push an event $e$ to the server (including the case when it is invoked by `exec`), we make additional invocations of `push`() so that events preceding $e$ in $\mathsf{ar}$ are pushed as well, in the order specified by $\mathsf{ar}$.

Our construction of $\mathcal{C}$ ensures that all events that should be pushed to the server according to the above rules before executing an event $e$ have already been executed.

We now define an order $Q$ in which the protocol executes the bodies of `exec`() functions. For the computation $\mathcal{C}$ to reproduce the history $\mathcal{H}$, the order $Q$ should include several relations. First, we must have $\mathsf{rt} \subseteq Q$, so that the protocol computation was consistent with the real-time order specified by $\mathcal{H}$ (note that this implies $\mathsf{so} \subseteq Q$). We must also have $\mathsf{vis} \subseteq Q$: to execute an operation, we should first execute all operations it is supposed to observe (rule 1 above). We must similarly have $\overline{\mathsf{ar}} \triangleq \mathsf{ar} \cap (\mathsf{Event} \times \mathsf{EPush}) \subseteq Q$: according to rule 2 above, if $e \in E \cap \mathsf{EPush}$, then executing $e$ will require pushing all its $\mathsf{ar}$-predecessors to the server, and the inclusion ensures that they have already been executed. Finally, we include one more relation into $Q$ that is more subtle. We define a relation $< \subseteq E \times E$ as follows:

$$
\begin{aligned}
e < f \iff{} & \exists e', g'.\, e' \in \mathsf{EPull} \wedge e \xrightarrow{\mathsf{so}?} e' \wedge g' \neq e' \wedge \neg(g' \xrightarrow{\mathsf{vis}} e') \wedge \\
& ((\exists g''.\, g' \xrightarrow{\mathsf{ar}?} g'' \xrightarrow{\mathsf{vis} \setminus \mathsf{so}} f) \vee (f \in \mathsf{EPush} \wedge g' \xrightarrow{\mathsf{ar}?} f)).
\end{aligned}
$$

## APPENDIX

The intuition behind this relation is as follows: according to the the above rules, executing $f$ would force us to push to the server some event $g'$ that will then be seen by by a $\mathsf{pull}$ event $e'$ following $e$ in a session; however, $e'$ is not supposed to see $g'$. Hence, in this case we need to execute $e$ before $f$ and we require $< \subseteq Q$. We now prove the following key property that allows us to construct the desired $Q$.

▶ **Lemma 9.** $\mathsf{rt} \cup \mathsf{vis} \cup \overline{\mathsf{ar}} \cup <$ *is acyclic.*

To prove this lemma, we establish several auxiliary results.

▶ **Proposition 10.** *For any* $\mathcal{A} = ((E, \mathsf{so}, \mathsf{rt}), \mathsf{vis}, \mathsf{ar}) \in \mathsf{ExecGSC}$ *and any* $S$ *such that* $S \cap \mathsf{rt}^{-1} = \emptyset$*, we have* $\mathsf{rt} \,;\, S \,;\, \mathsf{rt} \subseteq \mathsf{rt}$.

**Proof.** Fix an execution $\mathcal{A} = ((E, \mathsf{so}, \mathsf{rt}), \mathsf{vis}, \mathsf{ar}) \in \mathsf{ExecGSC}$ and assume

$$e_1 \xrightarrow{\mathsf{rt}} e_2 \xrightarrow{S} e_3 \xrightarrow{\mathsf{rt}} e_4.$$

Since $\mathsf{rt}$ is an interval order, either $e_1 \xrightarrow{\mathsf{rt}} e_4$ or $e_3 \xrightarrow{\mathsf{rt}} e_2$. However, the latter contradicts $S \cap \mathsf{rt}^{-1} = \emptyset$. Hence, we must have $e_1 \xrightarrow{\mathsf{rt}} e_4$, as desired. ◻

From Proposition 10 using ObservedAr and PushedAr we obtain

▶ **Corollary 11.** *For* $\mathcal{A} = ((E, \mathsf{so}, \mathsf{rt}), \mathsf{vis}, \mathsf{ar}) \in \mathsf{ExecGSC}$ *we have*

$$\mathsf{rt} \,;\, \mathsf{vis} \,;\, \mathsf{rt} \subseteq \mathsf{rt};$$
$$\mathsf{rt} \,;\, \overline{\mathsf{ar}} \,;\, \mathsf{rt} \subseteq \mathsf{rt}.$$

▶ **Proposition 12.** $<$ *is a strict partial order.*

**Proof.** We first prove that $<$ is transitive. Assume $e < f$ and $f < g$. Then for some $g_1', g_2'$ and $e', f' \in \mathsf{EPull}$ we have

$$e \xrightarrow{\mathsf{so}?} e' \wedge g_1' \neq e' \wedge \neg(g_1' \xrightarrow{\mathsf{vis}} e') \wedge ((\exists g_1''.\, g_1' \xrightarrow{\mathsf{ar}?} g_1'' \xrightarrow{\mathsf{vis} \setminus \mathsf{so}} f) \vee (f \in \mathsf{EPush} \wedge g_1' \xrightarrow{\mathsf{ar}?} f));$$

$$f \xrightarrow{\mathsf{so}?} f' \wedge g_2' \neq f' \wedge \neg(g_2' \xrightarrow{\mathsf{vis}} f') \wedge ((\exists g_2''.\, g_2' \xrightarrow{\mathsf{ar}?} g_2'' \xrightarrow{\mathsf{vis} \setminus \mathsf{so}} g) \vee (g \in \mathsf{EPush} \wedge g_2' \xrightarrow{\mathsf{ar}?} g)).$$

If for some $g_1''$ we have $g_1' \xrightarrow{\mathsf{ar}?} g_1'' \xrightarrow{\mathsf{vis} \setminus \mathsf{so}} f$, then $g_1' \xrightarrow{\mathsf{ar}?} g_1'' \xrightarrow{\mathsf{vis} \setminus \mathsf{so}} f'$. Since $\neg(g_2' \xrightarrow{\mathsf{vis}} f')$, by ObservedVis we must have $g_1'' \xrightarrow{\mathsf{ar}} g_2'$; then $g_1' \xrightarrow{\mathsf{ar}} g_2'$. If $f \in \mathsf{EPush}$ and $g_1' \xrightarrow{\mathsf{ar}?} f$, then $g_1' \xrightarrow{\mathsf{ar}?} f \xrightarrow{\mathsf{rt}?} f'$. Since $f' \in \mathsf{EPull}$ and $\neg(g_2' \xrightarrow{\mathsf{vis}} f')$, by PushedVis we again must have $g_1' \xrightarrow{\mathsf{ar}} g_2'$. This yields

$$e \xrightarrow{\mathsf{so}?} e' \wedge g_1' \neq e' \wedge \neg(g_1' \xrightarrow{\mathsf{vis}} e') \wedge ((\exists g_2''.\, g_1' \xrightarrow{\mathsf{ar}} g_2'' \xrightarrow{\mathsf{vis} \setminus \mathsf{so}} g) \vee (g \in \mathsf{EPush} \wedge g_1' \xrightarrow{\mathsf{ar}?} g)),$$

so that $e < g$.

We cannot have $e < e$, for in this case for some $e' \in \mathsf{EPull}$ we would have

$$\exists g'.\, e \xrightarrow{\mathsf{so}?} e' \wedge g' \neq e' \wedge \neg(g' \xrightarrow{\mathsf{vis}} e') \wedge ((\exists g''.\, g' \xrightarrow{\mathsf{ar}?} g'' \xrightarrow{\mathsf{vis} \setminus \mathsf{so}} e) \vee (e \in \mathsf{EPush} \wedge g' \xrightarrow{\mathsf{ar}?} e)).$$

Then

$$\exists g'.\, g' \neq e' \wedge \neg(g' \xrightarrow{\mathsf{vis}} e') \wedge ((\exists g''.\, g' \xrightarrow{\mathsf{ar}?} g'' \xrightarrow{\mathsf{vis} \setminus \mathsf{so}} e') \vee (e \in \mathsf{EPush} \wedge g' \xrightarrow{\mathsf{ar}?} e \xrightarrow{\mathsf{rt}?} e')).$$

contradicting ObservedVis or PushedVis. Hence, $<$ is a strict partial order. ◻

## APPENDIX

▶ **Proposition 13.** $< ; \overline{\mathsf{ar}} \subseteq <$.

**Proof.** Assume $e < f \xrightarrow{\mathsf{ar}} g$ and $g \in \mathsf{EPush}$. Then for some $g'$ and $e' \in \mathsf{EPull}$ we have

$$e \xrightarrow{\mathsf{so}?} e' \wedge g' \neq e' \wedge \neg(g' \xrightarrow{\mathsf{vis}} e') \wedge ((\exists g''.\, g' \xrightarrow{\mathsf{ar}?} g'' \xrightarrow{\mathsf{vis}\setminus\mathsf{so}} f) \vee (f \in \mathsf{EPush} \wedge g' \xrightarrow{\mathsf{ar}?} f));$$

Then $g' \xrightarrow{\mathsf{ar}} g$, so that

$$e \xrightarrow{\mathsf{so}?} e' \wedge g' \neq e' \wedge \neg(g' \xrightarrow{\mathsf{vis}} e') \wedge (g \in \mathsf{EPush} \wedge g' \xrightarrow{\mathsf{ar}} g).$$

This means $e < g$. ◻

▶ **Proposition 14.** $< ; (\mathsf{vis} \setminus \mathsf{so}) \subseteq <$.

**Proof.** Assume $e < f \xrightarrow{\mathsf{vis}\setminus\mathsf{so}} g$. Then

$$\exists e', g'.\, e' \in \mathsf{EPull} \wedge e \xrightarrow{\mathsf{so}?} e' \wedge g' \neq e' \wedge \neg(g' \xrightarrow{\mathsf{vis}} e') \wedge$$
$$((\exists g''.\, g' \xrightarrow{\mathsf{ar}?} g'' \xrightarrow{\mathsf{vis}\setminus\mathsf{so}} f \xrightarrow{\mathsf{vis}\setminus\mathsf{so}} g) \vee (f \in \mathsf{EPush} \wedge g' \xrightarrow{\mathsf{ar}?} f \xrightarrow{\mathsf{vis}\setminus\mathsf{so}} g)).$$

Hence,

$$\exists e', g'.\, e' \in \mathsf{EPull} \wedge e \xrightarrow{\mathsf{so}?} e' \wedge g' \neq e' \wedge \neg(g' \xrightarrow{\mathsf{vis}} e') \wedge g' \xrightarrow{\mathsf{ar}?} f \xrightarrow{\mathsf{vis}\setminus\mathsf{so}} g,$$

which implies $e < g$. ◻

▶ **Proposition 15.** $< ; \mathsf{rt} ; ((\mathsf{vis} \setminus \mathsf{so}) \cup \overline{\mathsf{ar}})? ; < \subseteq <$.

**Proof.** Assume

$$e_1 < e_2 \xrightarrow{\mathsf{rt};((\mathsf{vis}\setminus\mathsf{so})\cup\overline{\mathsf{ar}})?} e_3 < e_4.$$

Then for some $g'_1, g'_2$ and $e'_1, e'_3 \in \mathsf{EPull}$ we have

$$e_1 \xrightarrow{\mathsf{so}?} e'_1 \wedge g'_1 \neq e'_1 \wedge \neg(g'_1 \xrightarrow{\mathsf{vis}} e'_1) \wedge ((\exists g''_1.\, g'_1 \xrightarrow{\mathsf{ar}?} g''_1 \xrightarrow{\mathsf{vis}\setminus\mathsf{so}} e_2) \vee (e_2 \in \mathsf{EPush} \wedge g'_1 \xrightarrow{\mathsf{ar}?} e_2));$$

$$e_3 \xrightarrow{\mathsf{so}?} e'_3 \wedge g'_2 \neq e'_3 \wedge \neg(g'_2 \xrightarrow{\mathsf{vis}} e'_3) \wedge ((\exists g''_2.\, g'_2 \xrightarrow{\mathsf{ar}?} g''_2 \xrightarrow{\mathsf{vis}\setminus\mathsf{so}} e_4) \vee (e_4 \in \mathsf{EPush} \wedge g'_2 \xrightarrow{\mathsf{ar}?} e_4)).$$

Consider an arbitrary $g$ such that $g \xrightarrow{\mathsf{ar}?} g'_1$. We have

$$(\exists g''_1.\, g \xrightarrow{\mathsf{ar}?} g''_1 \xrightarrow{\mathsf{vis}\setminus\mathsf{so}} e_2 \xrightarrow{\mathsf{rt};((\mathsf{vis}\setminus\mathsf{so})\cup\overline{\mathsf{ar}})?;\mathsf{so}?} e'_3) \vee$$
$$(e_2 \in \mathsf{EPush} \wedge g \xrightarrow{\mathsf{ar}?} e_2 \xrightarrow{\mathsf{rt};((\mathsf{vis}\setminus\mathsf{so})\cup\overline{\mathsf{ar}})?;\mathsf{so}?} e'_3).$$

Hence, either

$$(\exists g''_1.\, g \xrightarrow{\mathsf{ar}?} g''_1 \xrightarrow{\mathsf{vis}\setminus\mathsf{so}} e_2 \xrightarrow{\mathsf{rt};((\mathsf{vis}\setminus\mathsf{so})\cup\overline{\mathsf{ar}});\mathsf{rt}?} e'_3) \vee$$
$$(e_2 \in \mathsf{EPush} \wedge g \xrightarrow{\mathsf{ar}?} e_2 \xrightarrow{\mathsf{rt};((\mathsf{vis}\setminus\mathsf{so})\cup\overline{\mathsf{ar}});\mathsf{rt}?} e'_3). \tag{3}$$

or

$$(\exists g''_1.\, g \xrightarrow{\mathsf{ar}?} g''_1 \xrightarrow{\mathsf{vis}\setminus\mathsf{so}} e_2 \xrightarrow{\mathsf{rt}} e'_3) \vee (e_2 \in \mathsf{EPush} \wedge g \xrightarrow{\mathsf{ar}?} e_2 \xrightarrow{\mathsf{rt}} e'_3). \tag{4}$$

If (3) holds, then by ObservedAr and PushedAr we have

$$g \xrightarrow{\mathsf{ar};((\mathsf{vis}\setminus\mathsf{so})\cup\overline{\mathsf{ar}});\mathsf{rt}?} e'_3.$$

Then by ObservedVis and PushedVis we get $g \xrightarrow{\mathsf{vis}} e'_3$. If (4) holds, then by ObservedVis and PushedVis we again get $g \xrightarrow{\mathsf{vis}} e'_3$. Thus, in any case we have $g \xrightarrow{\mathsf{vis}} e'_3$. Since $\neg(g'_2 \xrightarrow{\mathsf{vis}} e'_3)$, we hence must have $g'_1 \xrightarrow{\mathsf{ar}} g'_2$. Then

$$e_1 \xrightarrow{\mathsf{so}?} e'_1 \wedge g'_1 \neq e'_1 \wedge \neg(g'_1 \xrightarrow{\mathsf{vis}} e'_1) \wedge ((\exists g''_2.\, g'_1 \xrightarrow{\mathsf{ar}?} g''_2 \xrightarrow{\mathsf{vis}\setminus\mathsf{so}} e_4) \vee (e_4 \in \mathsf{EPush} \wedge g'_1 \xrightarrow{\mathsf{ar}?} e_4)),$$

so that $e_1 < e_4$, as required. ◻

## APPENDIX

▶ **Proposition 16.** *Let*

$$S = \mathsf{rt} \cup (\mathsf{vis} \setminus \mathsf{so}) \cup \overline{\mathsf{ar}} \cup (\mathsf{rt} \ ; (\mathsf{vis} \setminus \mathsf{so})) \cup ((\mathsf{vis} \setminus \mathsf{so}) \ ; \mathsf{rt}) \cup (\mathsf{rt} \ ; \overline{\mathsf{ar}}) \cup (\overline{\mathsf{ar}} \ ; \mathsf{rt}).$$

*Then*

$$(\mathsf{rt} \cup \mathsf{vis} \cup \overline{\mathsf{ar}} \cup <)^+ \subseteq S \cup < \cup (S \ ; <) \cup (< \ ; S) \cup (S \ ; < \ ; S)$$

*and $S$ is transitive.*

**Proof**. Since $\mathsf{vis} \subseteq \mathsf{ar}$, we have

$$\mathsf{vis} \ ; \overline{\mathsf{ar}} \subseteq \overline{\mathsf{ar}}.$$

Using ObservedVis and the fact that $\mathsf{so} \subseteq \mathsf{rt}$, we also get

$$\overline{\mathsf{ar}} \ ; \mathsf{vis} = \overline{\mathsf{ar}} \ ; (\mathsf{vis} \setminus \mathsf{so}) \cup \overline{\mathsf{ar}} \ ; \mathsf{so} \subseteq \mathsf{vis} \cup \overline{\mathsf{ar}} \ ; \mathsf{rt}.$$

Hence,

$$(\mathsf{vis} \cup \overline{\mathsf{ar}})^+ \subseteq \mathsf{vis} \cup \overline{\mathsf{ar}} \cup \overline{\mathsf{ar}} \ ; \mathsf{rt}. \tag{5}$$

Then

$$\begin{aligned}
& (\mathsf{rt} \cup \mathsf{vis} \cup \overline{\mathsf{ar}})^+ \\
= \ & (\mathsf{vis} \cup \overline{\mathsf{ar}})^+ \cup (\mathsf{vis} \cup \overline{\mathsf{ar}})^* \ ; (\mathsf{rt} \ ; (\mathsf{vis} \cup \overline{\mathsf{ar}})^*)^+ \\
\subseteq \ & \mathsf{vis} \cup \overline{\mathsf{ar}} \cup \overline{\mathsf{ar}} \ ; \mathsf{rt} \cup (\mathsf{vis} \cup \overline{\mathsf{ar}} \cup \overline{\mathsf{ar}} \ ; \mathsf{rt})? \ ; (\mathsf{rt} \ ; (\mathsf{vis} \cup \overline{\mathsf{ar}} \cup \overline{\mathsf{ar}} \ ; \mathsf{rt})?)^+ && \text{by (5)} \\
\subseteq \ & \mathsf{vis} \cup \overline{\mathsf{ar}} \cup \overline{\mathsf{ar}} \ ; \mathsf{rt} \cup (\mathsf{vis} \cup \overline{\mathsf{ar}} \cup \overline{\mathsf{ar}} \ ; \mathsf{rt})? \ ; (\mathsf{rt} \ ; (\mathsf{vis} \cup \overline{\mathsf{ar}})?)^+ && \text{by Corollary 11} \\
\subseteq \ & \mathsf{vis} \cup \overline{\mathsf{ar}} \cup \overline{\mathsf{ar}} \ ; \mathsf{rt} \cup (\mathsf{vis} \cup \overline{\mathsf{ar}} \cup \overline{\mathsf{ar}} \ ; \mathsf{rt})? \ ; \mathsf{rt} \ ; (\mathsf{vis} \cup \overline{\mathsf{ar}})? && \text{by Corollary 11} \\
= \ & \mathsf{vis} \cup \overline{\mathsf{ar}} \cup \overline{\mathsf{ar}} \ ; \mathsf{rt} \cup \mathsf{rt} \cup \mathsf{vis} \ ; \mathsf{rt} \cup \mathsf{rt} \ ; \mathsf{vis} \cup \mathsf{rt} \ ; \overline{\mathsf{ar}} \cup \\
& (\mathsf{vis} \cup \overline{\mathsf{ar}} \cup \overline{\mathsf{ar}} \ ; \mathsf{rt}) \ ; \mathsf{rt} \ ; (\mathsf{vis} \cup \overline{\mathsf{ar}}) \\
= \ & \mathsf{vis} \cup \overline{\mathsf{ar}} \cup \overline{\mathsf{ar}} \ ; \mathsf{rt} \cup \mathsf{rt} \cup \mathsf{vis} \ ; \mathsf{rt} \cup \mathsf{rt} \ ; \mathsf{vis} \cup \mathsf{rt} \ ; \overline{\mathsf{ar}} \cup \\
& (\mathsf{vis} \cup \overline{\mathsf{ar}}) \ ; \mathsf{rt} \ ; (\mathsf{vis} \cup \overline{\mathsf{ar}}) \\
\subseteq \ & (\mathsf{vis} \setminus \mathsf{so}) \cup \overline{\mathsf{ar}} \cup \overline{\mathsf{ar}} \ ; \mathsf{rt} \cup \mathsf{rt} \cup (\mathsf{vis} \setminus \mathsf{so}) \ ; \mathsf{rt} \cup \mathsf{rt} \ ; (\mathsf{vis} \setminus \mathsf{so}) \cup \mathsf{rt} \ ; \overline{\mathsf{ar}} \cup && \text{since } \mathsf{so} \subseteq \mathsf{rt} \\
& ((\mathsf{vis} \setminus \mathsf{so}) \cup \overline{\mathsf{ar}}) \ ; \mathsf{rt} \ ; ((\mathsf{vis} \setminus \mathsf{so}) \cup \overline{\mathsf{ar}}) \\
= \ & (\mathsf{vis} \setminus \mathsf{so}) \cup \overline{\mathsf{ar}} \cup \overline{\mathsf{ar}} \ ; \mathsf{rt} \cup \mathsf{rt} \cup (\mathsf{vis} \setminus \mathsf{so}) \ ; \mathsf{rt} \cup \mathsf{rt} \ ; (\mathsf{vis} \setminus \mathsf{so}) \cup \mathsf{rt} \ ; \overline{\mathsf{ar}} \cup \\
& (\mathsf{vis} \setminus \mathsf{so}) \ ; \mathsf{rt} \ ; (\mathsf{vis} \setminus \mathsf{so}) \cup (\mathsf{vis} \setminus \mathsf{so}) \ ; \mathsf{rt} \ ; \overline{\mathsf{ar}} \cup \\
& \overline{\mathsf{ar}} \ ; \mathsf{rt} \ ; (\mathsf{vis} \setminus \mathsf{so}) \cup \overline{\mathsf{ar}} \ ; \mathsf{rt} \ ; \overline{\mathsf{ar}} \\
\subseteq \ & (\mathsf{vis} \setminus \mathsf{so}) \cup \overline{\mathsf{ar}} \cup \overline{\mathsf{ar}} \ ; \mathsf{rt} \cup \mathsf{rt} \cup (\mathsf{vis} \setminus \mathsf{so}) \ ; \mathsf{rt} \cup \mathsf{rt} \ ; (\mathsf{vis} \setminus \mathsf{so}) \cup \mathsf{rt} \ ; \overline{\mathsf{ar}} \cup && \text{by ObservedAr} \\
& \mathsf{ar} \ ; (\mathsf{vis} \setminus \mathsf{so}) \cup \mathsf{ar} \ ; \overline{\mathsf{ar}} \cup \overline{\mathsf{ar}} \ ; \mathsf{ar} \ ; (\mathsf{vis} \setminus \mathsf{so}) \cup \overline{\mathsf{ar}} \ ; \mathsf{ar} \ ; \overline{\mathsf{ar}} && \text{and PushedAr} \\
= \ & (\mathsf{vis} \setminus \mathsf{so}) \cup \overline{\mathsf{ar}} \cup \overline{\mathsf{ar}} \ ; \mathsf{rt} \cup \mathsf{rt} \cup (\mathsf{vis} \setminus \mathsf{so}) \ ; \mathsf{rt} \cup \mathsf{rt} \ ; (\mathsf{vis} \setminus \mathsf{so}) \cup \mathsf{rt} \ ; \overline{\mathsf{ar}} \cup \\
& \mathsf{ar} \ ; (\mathsf{vis} \setminus \mathsf{so}) \\
\subseteq \ & (\mathsf{vis} \setminus \mathsf{so}) \cup \overline{\mathsf{ar}} \cup \overline{\mathsf{ar}} \ ; \mathsf{rt} \cup \mathsf{rt} \cup (\mathsf{vis} \setminus \mathsf{so}) \ ; \mathsf{rt} \cup \mathsf{rt} \ ; (\mathsf{vis} \setminus \mathsf{so}) \cup \mathsf{rt} \ ; \overline{\mathsf{ar}} \cup && \text{by ObservedVis} \\
& \mathsf{vis} \\
= \ & \mathsf{rt} \cup (\mathsf{vis} \setminus \mathsf{so}) \cup \overline{\mathsf{ar}} \cup (\mathsf{rt} \ ; (\mathsf{vis} \setminus \mathsf{so})) \cup \\
& ((\mathsf{vis} \setminus \mathsf{so}) \ ; \mathsf{rt}) \cup (\mathsf{rt} \ ; \overline{\mathsf{ar}}) \cup (\overline{\mathsf{ar}} \ ; \mathsf{rt}) \\
= \ & S
\end{aligned}$$

Additionally, by Propositions 12-15 we get $< \ ; S \ ; < \ \subseteq \ <$. Then we obtain the desired inclusion from this property, Proposition 12 and the above inclusion. Finally, $S$ is transitive because

$$S \ ; S \subseteq (\mathsf{rt} \cup \mathsf{vis} \cup \overline{\mathsf{ar}})^+ \subseteq S.$$

□

▶ **Proposition 17.** $< \ ; \mathsf{rt} \ ; ((\mathsf{vis} \setminus \mathsf{so}) \cup \overline{\mathsf{ar}})?$ *is irreflexive.*

**Proof.** Assume $e < f \xrightarrow{\mathsf{rt};((\mathsf{vis}\setminus\mathsf{so})\cup\overline{\mathsf{ar}})?} e$. Then for some $e' \in \mathsf{EPull}$ we have

$$\exists g'.\, e \xrightarrow{\mathsf{so}?} e' \wedge g' \neq e' \wedge \neg(g' \xrightarrow{\mathsf{vis}} e') \wedge \\ ((\exists g''.\, g' \xrightarrow{\mathsf{ar}} g'' \xrightarrow{\mathsf{vis}\setminus\mathsf{so}} f \xrightarrow{\mathsf{rt};((\mathsf{vis}\setminus\mathsf{so})\cup\overline{\mathsf{ar}})?} e) \vee (f \in \mathsf{EPush} \wedge g' \xrightarrow{\mathsf{ar}?} f \xrightarrow{\mathsf{rt};((\mathsf{vis}\setminus\mathsf{so})\cup\overline{\mathsf{ar}})?} e)).s$$

Then

$$\exists g'.\, g' \neq e' \wedge \neg(g' \xrightarrow{\mathsf{vis}} e') \wedge \\ ((\exists g''.\, g' \xrightarrow{\mathsf{ar}} g'' \xrightarrow{\mathsf{vis}\setminus\mathsf{so}} f \xrightarrow{\mathsf{rt};((\mathsf{vis}\setminus\mathsf{so})\cup\overline{\mathsf{ar}})?;\mathsf{so}?} e') \vee (f \in \mathsf{EPush} \wedge g' \xrightarrow{\mathsf{ar}?} f \xrightarrow{\mathsf{rt};((\mathsf{vis}\setminus\mathsf{so})\cup\overline{\mathsf{ar}})?;\mathsf{so}?} e')).$$

We have that either

$$\exists g'.\, g' \neq e' \wedge \neg(g' \xrightarrow{\mathsf{vis}} e') \wedge \\ ((\exists g''.\, g' \xrightarrow{\mathsf{ar}} g'' \xrightarrow{\mathsf{vis}\setminus\mathsf{so}} f \xrightarrow{\mathsf{rt};((\mathsf{vis}\setminus\mathsf{so})\cup\overline{\mathsf{ar}});\mathsf{rt}?} e') \vee (f \in \mathsf{EPush} \wedge g' \xrightarrow{\mathsf{ar}?} f \xrightarrow{\mathsf{rt};((\mathsf{vis}\setminus\mathsf{so})\cup\overline{\mathsf{ar}});\mathsf{rt}?} e')). \tag{6}$$

or

$$\exists g'.\, g' \neq e' \wedge \neg(g' \xrightarrow{\mathsf{vis}} e') \wedge ((\exists g''.\, g' \xrightarrow{\mathsf{ar}} g'' \xrightarrow{\mathsf{vis}\setminus\mathsf{so}} f \xrightarrow{\mathsf{rt}} e') \vee (f \in \mathsf{EPush} \wedge g' \xrightarrow{\mathsf{ar}?} f \xrightarrow{\mathsf{rt}} e')). \tag{7}$$

If (6) holds, then by ObservedAr and PushedAr we get

$$\exists g'.\, g' \neq e' \wedge \neg(g' \xrightarrow{\mathsf{vis}} e') \wedge g' \xrightarrow{\mathsf{ar};((\mathsf{vis}\setminus\mathsf{so})\cup\overline{\mathsf{ar}});\mathsf{rt}?} e',$$

which contradicts ObservedVis or PushedVis. Similarly, (7) contradicts either ObservedVis or PushedVis. ◻

**Proof of Lemma 9.** Assume that $\mathsf{rt} \cup \mathsf{vis} \cup \overline{\mathsf{ar}} \cup <$ contains a cycle. Since this relation is irreflexive, by Proposition 16 for some $e$ we have

$$(e,e) \in ((\mathsf{vis}\setminus\mathsf{so}) \,;\, \mathsf{rt}) \cup (\overline{\mathsf{ar}} \,;\, \mathsf{rt}) \cup \\ (< \,;\, (\mathsf{rt} \cup (\mathsf{vis}\setminus\mathsf{so}) \cup \overline{\mathsf{ar}} \cup (\mathsf{rt} \,;\, (\mathsf{vis}\setminus\mathsf{so})) \cup ((\mathsf{vis}\setminus\mathsf{so}) \,;\, \mathsf{rt}) \cup (\mathsf{rt} \,;\, \overline{\mathsf{ar}}) \cup (\overline{\mathsf{ar}} \,;\, \mathsf{rt}))).$$

Then by Propositions 13 and 14,

$$(e,e) \in ((\mathsf{vis}\setminus\mathsf{so}) \,;\, \mathsf{rt}) \cup (\overline{\mathsf{ar}} \,;\, \mathsf{rt}) \cup (< \,;\, \mathsf{rt}) \cup (< \,;\, \mathsf{rt} \,;\, (\mathsf{vis}\setminus\mathsf{so})) \cup (< \,;\, \mathsf{rt} \,;\, \overline{\mathsf{ar}}).$$

ObservedAr implies that $((\mathsf{vis}\setminus\mathsf{so}) \,;\, \mathsf{rt})$ is irreflexive, and PushedAr implies that so is $(\overline{\mathsf{ar}} \,;\, \mathsf{rt})$. Hence,

$$(e,e) \in (< \,;\, \mathsf{rt}) \cup (< \,;\, \mathsf{rt} \,;\, (\mathsf{vis}\setminus\mathsf{so})) \cup (< \,;\, \mathsf{rt} \,;\, \overline{\mathsf{ar}}).$$

But this contradicts Proposition 17. This contradiction shows that $< \cup\, \mathsf{rt} \cup \mathsf{vis} \cup \overline{\mathsf{ar}}$ must be acyclic. ◻

▶ **Proposition 18.** $<$ *is prefix-finite.*

**Proof.** Fix $f \in E$; we show that there are only finitely many $e$ such that $e < f$. Since $\mathsf{vis}$ and $\mathsf{ar}$ are prefix-finite, there are only finitely many $g'$ such that

$$(\exists g''.\, g' \xrightarrow{\mathsf{ar}?} g'' \xrightarrow{\mathsf{vis}\setminus\mathsf{so}} f) \vee (f \in \mathsf{EPush} \wedge g' \xrightarrow{\mathsf{ar}?} f).$$

By Eventual, there are only finitely many $e'$ such that for some $g'$ satisfying the above property we have $\neg(g' \xrightarrow{\mathsf{vis}} e')$. Since $\mathsf{so}$ is prefix-finite and , this implies the required. ◻

## APPENDIX

▶ **Lemma 19.** $\mathsf{rt} \cup \mathsf{vis} \cup \overline{\mathsf{ar}} \cup <$ *is prefix-finite.*

**Proof.** Follows from Propositions 16 and 18 and the prefix-finiteness of rt, vis and ar. ◻

▶ **Lemma 20.** *For any history $\mathcal{H}$ and a prefix-finite acyclic relation $S \subseteq E_{\mathcal{H}} \times E_{\mathcal{H}}$ such that $\mathsf{so}_{\mathcal{H}} \subseteq S$, there exists a total prefix-finite order on $E_{\mathcal{H}}$ containing $S$.*

**Proof.** Let $\mathcal{H} = (E, \mathsf{so}, \mathsf{rt})$. Recall that so is a union of total orders over a finite number of disjoint subsets of $E$, which represent different sessions. We use this fact to construct the desired order $S' \supseteq S$ by scheduling events according to a certain strategy. The order is constructed inductively, so that its prefix-finiteness holds by construction. At every step of the construction we have a frontier $E_0$ of events from different sessions that can be scheduled next. We use a round-robin scheduling strategy among the sessions with an additional constraint that we skip any event $e \in E_0$ if there is another event $f \in E_0$ such that $(f, e) \in S^+$. Since $S$ is acyclic, there is always at least one event to be scheduled. The relation $S'$ constructed in the above way is prefix-finite by construction.

The order $S'$ is total on $E$. Indeed, assume the contrary: there is an event $e \in E$ that gets ignored forever by the scheduling strategy. Without a loss of generality, we assume that $e$ is the minimal such event in its session. Then at some point all its so-predecessors are scheduled, and after this, every time the round-robin scheduler passes the session of $e$, there is another event $f$ on the frontier that is scheduled instead of $e$; then $(f, e) \in S^+$. Hence, we must have infinitely many such events $f$, contradicting the prefix-finiteness of $S$.

Finally, we show $S \subseteq S'$. Assume the contrary: for some $e, f$ we have $(e, f) \in S$, but $(f, e) \in S'$. Since $\mathsf{so} \subseteq S$ and $\mathsf{so} \subseteq S'$, the events $e$ and $f$ must be in different sessions. Let $e'$ be the event on the frontier of the session of $e$ at the time $f$ was scheduled; then $(e', e) \in \mathsf{so}?$. Since $(e, f) \in S$ and $\mathsf{so} \subseteq S$, we have $(e', f) \in S^+$. But then we could not have scheduled $f$ according to our strategy. ◻

By Lemmas 9, 19 and 20, there exists a prefix-finite total order $Q$ on events in $E$ containing $\mathsf{rt} \cup \mathsf{vis} \cup \overline{\mathsf{ar}} \cup <$. We use this order to determine the order of executing the bodies of exec() functions. To determine the order of executing calls to and returns from exec(), we use the following adjustment of a classical result about interval orders [13].

▶ **Lemma 21.** *For a history $\mathcal{H} = (E, \mathsf{so}, \mathsf{rt})$, let $E'$ be a set consisting of special events of the form $(\iota', \mathsf{call}(e))$ and $(\iota'', \mathsf{return}(e))$, one pair for each event $e \in E$. Then there exists a total prefix-finite order $Q' \subseteq E' \times E'$ such that*

$$\forall e \in E.\, (_, \mathsf{call}(e)) \xrightarrow{Q'} (_, \mathsf{return}(e)); \qquad (8)$$

$$\forall e, f \in E.\, e \xrightarrow{\mathsf{rt}} f \iff (_, \mathsf{return}(e)) \xrightarrow{Q'} (_, \mathsf{call}(f)). \qquad (9)$$

**Proof.** Similarly to Lemma 20, we construct the desired order $Q'$ inductively, using the fact that $E$ is partitioned by so into finitely many sessions. At every step, the set $E$ is partitioned into the set $E'$ of events for which we have added calls to $Q'$, and the set $E''$ of those events for which we have not; the elements of $E''$ minimal in so form the frontier $E_0$. We first consider events $e \in E'$ that lack a matching return$(e)$ and append return$(e)$ to $Q'$ for all such events $e$ such that $\forall f \in E''.\, e \xrightarrow{\mathsf{rt}} f$. We then choose the next event from $E_0$ to add a call for by using a round-robin scheduling strategy among the sessions. This is subject to an additional restriction that we skip any event $e \in E_0$ if there is either another event $f \in E_0$ such that $f \xrightarrow{\mathsf{rt}} e$, or there is an event $f \in E'$ without a return$(f)$ in $Q'$ such that $f \xrightarrow{\mathsf{rt}} e$.

We now argue that the above scheduling strategy never gets stuck: if the set $E_0$ is non-empty, then we can choose an event to process. Assume the contrary: $E_0$ is non-empty, but all its events get discarded by scheduling strategy. Let $E'_0$ be the subset of $E_0$ consisting of $\mathsf{rt}$-minimal events. Since $\mathsf{rt}$ is a strict partial order, $E'_0$ is non-empty. Let $E'_0 = \{e_1, \dots, e_n\}$, $n \geq 1$. Since none of the events in $E'_0$ can be chosen for processing, for every $e_i \in E'_0$ there exists $f_i \in E'$ without a $\mathsf{return}(f_i)$ in $Q'$ such that $f \xrightarrow{\mathsf{rt}} e$. Let $F = \{f_1, \dots, f_n\}$. Assume there are $i, j$ such that $i \neq j$ and $f_i \neq f_j$. Then $f_i \xrightarrow{\mathsf{rt}} e_i$ and $f_j \xrightarrow{\mathsf{rt}} e_j$. Since $\mathsf{rt}$ is an interval order, either $f_i \xrightarrow{\mathsf{rt}} e_j$ or $f_j \xrightarrow{\mathsf{rt}} e_i$. In the former case, we can replace $f_j$ by $f_i$ in $F$, and in the latter case, $f_i$ by $f_j$. Continuing this process, we can find a single $f_k$ such that $\forall e \in E'_0.\, f_k \xrightarrow{\mathsf{rt}} e$. But in this case, our construction would append a $\mathsf{return}(f_k)$ to $Q'$, contradicting our assumption about $f_k$. Thus, the scheduling strategy always makes progress.

Since we have $\forall e \in E.\, |\{f \in E \mid \neg(e \xrightarrow{\mathsf{rt}} f)\}| < \infty$, for every $\mathsf{call}(e)$ added to $Q'$ we will eventually add a matching $\mathsf{return}(e)$. The totality of the $Q'$ resulting from our construction and its prefix-finiteness are justified as in Lemma 20. Property (8) holds by the construction of $Q'$.

Consider any $e, f$ such that $(_, \mathsf{return}(e)) \xrightarrow{Q'} (_, \mathsf{call}(f))$. Then by the rule for adding $\mathsf{return}$ events to $Q'$, we must have $e \xrightarrow{\mathsf{rt}} f$. Hence, the $\Longleftarrow$ direction of (9) holds.

Now assume that $e \xrightarrow{\mathsf{rt}} f$. Since $\mathsf{so} \subseteq \mathsf{rt}$, our scheduling strategy always selects an $\mathsf{rt}$-minimal event among the unprocessed ones. Hence, $(_, \mathsf{call}(e)) \xrightarrow{Q'} (_, \mathsf{call}(f))$. But then we cannot have $(_, \mathsf{call}(f)) \xrightarrow{Q'} (_, \mathsf{return}(e))$: in this case at the moment of adding $\mathsf{call}(f)$ to $Q'$ we would have $\mathsf{call}(e)$ in $Q'$, but not yet $\mathsf{return}(e)$; since $e \xrightarrow{\mathsf{rt}} f$, this would contradict the scheduling strategy. Hence, the $\Longrightarrow$ direction of (9) holds as well. ◻

Let $E'$ and $Q'$ be the set and the relation constructed in the above lemma for the history $\mathcal{H}$. Let $E'' = E \uplus E'$ and

$$Q'' = \{(((_, \mathsf{call}(e))), e), (e, \mathsf{return}(e)) \mid e \in E\}.$$

▶ **Lemma 22.** *$Q \cup Q' \cup Q''$ is acyclic and prefix-finite.*

**Proof.** It is easy to see that

$$(Q' \cup Q'')^+ \subseteq Q' \cup Q'' \cup (Q' \,;\, Q'') \cup (Q'' \,;\, Q') \cup (Q'' \,;\, Q' \,;\, Q'').$$

Hence, $Q' \cup Q''$ is acyclic and prefix-finite.

We now show that

$$Q'' \,;\, Q \,;\, Q'' \subseteq Q'. \tag{10}$$

Indeed, assume that for some $e, f \in E$ we have

$$(_, \mathsf{call}(e)) \xrightarrow{Q''} e \xrightarrow{Q} f \xrightarrow{Q''} (_, \mathsf{return}(f)).$$

We cannot have $(_, \mathsf{return}(f)) \xrightarrow{Q'} (_, \mathsf{call}(e))$, for in this case by the definition of $Q'$ we would have $f \xrightarrow{\mathsf{rt}} e$, contradicting the definition of $Q$. Hence, we must have $(_, \mathsf{call}(e)) \xrightarrow{Q'} (_, \mathsf{return}(f))$, as required.

From (10) it follows that

$$(Q \cup Q' \cup Q'')^+ \subseteq (Q' \cup Q'')^+ \cup ((Q' \cup Q'')^+? \,;\, Q) \cup (Q \,;\, (Q' \cup Q'')^+?) \cup (Q \,;\, (Q' \cup Q'')^+ \,;\, Q).$$

This and (10) implies that $Q \cup Q' \cup Q''$ is acyclic. Since $Q$ and $Q' \cup Q''$ are prefix-finite, this also implies that $Q \cup Q' \cup Q''$ is prefix-finite. ◻

By Lemma 22 and an easy adjustment of Lemma 20, we get that there exists a total prefix-finite order $Q''' \subseteq E'' \times E''$. To construct the desired computation $\mathcal{C}$ of the protocol, we execute calls to, returns from and bodies of exec() functions generating the events in $E$ in the order determined by $Q'''$. Since $Q$ is prefix-finite, every all transitions corresponding to $E''$ will be executed.

To prove that the computation $\mathcal{C}$ constructed in this way is indeed valid computation of the protocol, we first need to show that all events we require pushed to the server when executing any event $e$ according to rules 1 and 2 above have already executed before $e$. Indeed, consider such an event $f$. Then

$$(\exists g.\, f \xrightarrow{\mathsf{ar}?} g \xrightarrow{\mathsf{vis}\setminus\mathsf{so}} e) \vee (e \in \mathsf{EPush} \wedge f \xrightarrow{\mathsf{ar}} e).$$

Since $\mathcal{A}$ satisfies ObservedVis, this implies $(f, e) \in \mathsf{vis} \cup \overline{\mathsf{ar}} \subseteq Q$, so that $f$ must have been executed before $e$, as required.

We next prove that $\mathcal{C}$ satisfies the following invariant relating any of its prefixes $\mathcal{C}'$ with the abstract execution $\mathcal{A}$. First, the history extracted from $\mathcal{C}'$ as in §3 is equal to $\mathcal{H}$ projected onto the events executed in $\mathcal{C}'$. This implies that the history extracted from $\mathcal{C}$ is exactly $\mathcal{H}$, as desired. Further, from $\mathcal{C}'$ we can extract the following relations $\mathsf{vis}'$ and $\mathsf{ar}'$. The relation $\mathsf{vis}'$ is extracted in the same way as visibility in the proof of soundness: we have $e \xrightarrow{\mathsf{vis}'} f$ if the operation of $f$ was evaluated in $\mathcal{C}'$ on a log incorporating $e$. Unlike $\mathsf{ar}$, the relation $\mathsf{ar}'$ is partial. First, it orders events according to their order in *server_log*. It also orders events in the *pending* log of each client according to their order in this log and after all events in *server_log*. Our invariant requires that $\mathsf{vis}'$ be equal to the projection of $\mathsf{vis}$ to events executed in $\mathcal{C}'$ and $\mathsf{ar}'$ be a subset of the similar projection of $\mathsf{ar}$.

We prove the above invariant by induction on the length of the prefix of $\mathcal{C}$. Assume the invariant holds of a prefix $\mathcal{C}'$ and consider the next prefix $\mathcal{C}''$, obtained by executing an event $e$. Since $\mathsf{so} \subseteq Q$, the per-client order of events in $\mathcal{C}''$ follows $\mathsf{so}$. Let $\mathsf{vis}'$ and $\mathsf{ar}'$ be the relations extracted from $\mathcal{C}''$ as required by the invariant. The relationship between $\mathsf{ar}'$ and $\mathsf{ar}$ required by the invariant follows from rule 2 above. From rule 1 we also get that $\mathsf{vis}'$ is a superset of the projection of $\mathsf{vis}$ to the events in $\mathcal{C}''$. We next prove that the converse is also true. Since $\mathcal{A}$ satisfies RetVal, this implies that the value returned by the invocation of exec() that generated $e$ is equal to $\mathsf{rval}(e)$ (note that $\mathsf{ar}'$ is total on all events visible to the event $e$). We hence establish that the invariant holds of $\mathcal{C}''$.

Consider $e$ and $f$ such that $f \xrightarrow{\mathsf{vis}'} e$. Then $e \neq f$. Assume that $\neg(f \xrightarrow{\mathsf{vis}} e)$. Then by RYW, $f$ must be executed by a different client from $e$. Since $f \xrightarrow{\mathsf{vis}'} e$, the client of $e$ had pulled $f$ from the server by the time it executed $e$. Consider the event $e'$ by the client of $e$ such that $f$ was pulled by the exec() function that generated $e'$ or right before it. Then $e' \xrightarrow{\mathsf{so}?} e$. We cannot have $f \xrightarrow{\mathsf{vis}} e'$, for in this case by MonotonicView we would have $f \xrightarrow{\mathsf{vis}} e$, contradicting our assumption. But then by rule 2 above $f$ could only be pulled by the client executing $e'$ because $e' \in \mathsf{EPull}$ and $f$ was on the server when $e'$ was executed. Consider now the point in the computation when $f$ was pushed to the server. This happened because it was required by rules 1 or 2 for the execution of some event $f'$. Then

$$(\exists e_1.\, f \xrightarrow{\mathsf{ar}?} e_1 \xrightarrow{\mathsf{vis}\setminus\mathsf{so}} f') \vee (f' \in \mathsf{EPush} \wedge f \xrightarrow{\mathsf{ar}?} f').$$

At the point in the computation when $f'$ was executed, $e'$ has not yet been executed. Let $e''$ be the event by the client of $e'$ that was pending to be executed. Then $e'' \xrightarrow{\mathsf{so}?} e'$. All of the above implies $e'' < f'$, so that $(e'', f') \in Q$. But this means we could not execute $f'$ before $e''$. This contradiction shows that we must have $f \xrightarrow{\mathsf{vis}} e$, as desired.

The computation $\mathcal{C}$ is consistent with the GSC protocol in Figure 2, but may not satisfy the fairness constraints we required in §3. If the computation is finite, then by appending to it additional invocations of `push`() and `pull`() we can ensure that the fairness constraints are satisfied. Assume now that the computation is infinite, i.e., so is $E$.

By EVENTUAL, any event $e \in E$ cannot be invisible to infinitely many events from $E$. Hence, the only case when an event $e$ may not be pushed to the server in $\mathcal{C}$ is when the client executing $e$ is the only one that executes infinitely many operations in $\mathcal{A}$. In this case we can ensure that every event is pushed to the server by adding invocations of `push`() to $\mathcal{C}$ after clients that execute finitely many events have finished executing these events. This transformation produces a valid computation $\mathcal{C}_1$.

From EVENTUAL it also follows that in $\mathcal{C}_1$ if a client does not pull some event $e$, then either this client executes only finitely many events, or it executes infinitely many events, but $e$ is executed by the same client. In the former case, by adding invocations of `pull`() at this client in $\mathcal{C}_1$ after it finishes executing all its events we can ensure that this client eventually pulls all events. In the latter case, the server log must have an infinite suffix consisting only of events by the client in question. By adding invocations of `pull`() at this client we can again ensure that the client pulls all events without changing the validity of the computation. The computation $\mathcal{C}_2$ we obtain using the above transformations satisfies all the fairness constraints and, hence, is the desired one. □

# C Additional Material on Relationships between Consistency Models

## C.1 Proof of Theorem 1

We complete the proof given in §7 by establishing that GSP and TSO are equivalent modulo real-time order:

$$\forall E.\, \forall \mathsf{so}.\, E \cap (\mathsf{EPush} \cup \mathsf{EPull}) = \emptyset \implies ((\exists \mathsf{rt}.\, (E, \mathsf{so}, \mathsf{rt}) \in \mathsf{HistGSC}) \iff (\exists \mathsf{rt}''.\, (\mathsf{mkPull}(E), \mathsf{mkPull}(\mathsf{so}), \mathsf{rt}'') \in \mathsf{HistGSC})).$$

Consider $E$ and $\mathsf{so}$ such that $E \cap (\mathsf{EPush} \cup \mathsf{EPull}) = \emptyset$. It is easy to see that

$$\forall \mathsf{rt}.\, (\mathsf{mkPull}(E), \mathsf{mkPull}(\mathsf{so}), \mathsf{mkPull}(\mathsf{rt})) \in \mathsf{HistGSC} \implies (E, \mathsf{so}, \mathsf{rt}) \in \mathsf{HistGSC},$$

since erasing fences from events does not invalidate any axioms.

Hence, it remains to prove that

$$(\exists \mathsf{rt}.\, (E, \mathsf{so}, \mathsf{rt}) \in \mathsf{HistGSC}) \implies (\exists \mathsf{rt}''.\, (\mathsf{mkPull}(E), \mathsf{mkPull}(\mathsf{so}), \mathsf{rt}'') \in \mathsf{HistGSC}).$$

To this end, assume $\mathsf{rt}$ such that $(E, \mathsf{so}, \mathsf{rt}) \in \mathsf{HistGSC}$. Then for some $\mathsf{vis}$ and $\mathsf{ar}$ we have

$$\mathcal{A} = ((E, \mathsf{so}, \mathsf{rt}), \mathsf{vis}, \mathsf{ar}) \in \mathsf{ExecGSC}.$$

Let

$$E'' = \mathsf{mkPull}(E); \quad \mathsf{so}'' = \mathsf{mkPull}(\mathsf{so}); \quad \mathsf{vis}'' = \mathsf{mkPull}(\mathsf{vis}); \quad \mathsf{ar}'' = \mathsf{mkPull}(\mathsf{ar}).$$

Let $\mathsf{rt}''$ be any prefix-finite total order on $E''$ that contains $\mathsf{vis}'' \cup \mathsf{mkPull}(<)$; such an order exists by Lemmas 9, 19 and 20. Then $\mathcal{A}'' = ((E'', \mathsf{so}'', \mathsf{rt}''), \mathsf{vis}'', \mathsf{ar}'')$ is an abstract execution. We now show that $\mathcal{A}''$ satisfies all GSC axioms. The execution $\mathcal{A}''$ satisfies RETVAL, RYW, MONOTONICVIEW and EVENTUAL because so does $\mathcal{A}$. It satisfies PUSHEDVIS and PUSHEDAR because $E'' \cap \mathsf{EPush} = \emptyset$.

We next show that $\mathcal{A}''$ satisfies ObservedVis. We have $\mathsf{ar}'' \,;\, (\mathsf{vis}'' \setminus \mathsf{so}'') \subseteq \mathsf{vis}''$ because $\mathcal{A}$ satisfies ObservedVis. It thus remains to show that

$$\mathsf{ar}''? \,;\, (\mathsf{vis}'' \setminus \mathsf{so}'') \,;\, (\mathsf{rt}'' \cap (\mathsf{Event} \times \mathsf{EPull})) \subseteq \mathsf{vis}''.$$

Consider $g', g'', f, e' \in E''$ such that

$$g' \xrightarrow{\mathsf{ar}''?} g'' \xrightarrow{\mathsf{vis}''\setminus\mathsf{so}''} f \xrightarrow{\mathsf{rt}''} e'$$

and assume $(g', e') \notin \mathsf{vis}''$. Since $\mathcal{A}$ satisfies ObservedVis, we have $(g', f) \in \mathsf{vis}''$. Since $\mathsf{vis}'' \subseteq \mathsf{rt}''$, this implies $g' \neq e'$. But then $(e', f) \in \mathsf{mkPull}(<)$, contradicting the definition of $\mathsf{rt}''$. Hence, $\mathcal{A}''$ satisfies ObservedVis.

Finally, we show that $\mathcal{A}''$ satisfies ObservedAr. Consider $e, f, g \in E''$ such that

$$e \xrightarrow{\mathsf{vis}''\setminus\mathsf{so}''} f \xrightarrow{\mathsf{rt}''} g.$$

We have $e \neq g$ by the construction of $\mathsf{rt}''$. Hence, we must have either $e \xrightarrow{\mathsf{ar}''} g$ or $g \xrightarrow{\mathsf{ar}''} e$. Assume the latter, so that

$$g \xrightarrow{\mathsf{ar}''} e \xrightarrow{\mathsf{vis}''\setminus\mathsf{so}''} f \xrightarrow{\mathsf{rt}''} g.$$

Since $\mathcal{A}''$ satisfies ObservedVis, this implies

$$g \xrightarrow{\mathsf{vis}''} f \xrightarrow{\mathsf{rt}''} g,$$

which contradicts the definition of $\mathsf{rt}''$. Hence, we must have $e \xrightarrow{\mathsf{ar}''} g$, so that $\mathcal{A}''$ satisfies ObservedAr.

We have thus shown that $(E'', \mathsf{so}'', \mathsf{rt}'') \in \mathsf{HistGSC}$. ◻

## C.2 Correspondence with Linearizability

In our framework we can define the set of histories allowed by linearizability as follows:

$$\mathsf{HistLIN}_1 = \{\mathcal{H} \mid \mathcal{H} \in \mathsf{HistGSC} \wedge E_{\mathcal{H}} \subseteq \mathsf{EPush} \cap \mathsf{EPull}\}.$$

We now show that this coincides with the standard definition [15]. A *linearization* of a history $\mathcal{H}$ is a pair $(\mathcal{H}, \mathsf{lin})$, where $\mathsf{lin}$ is a prefix-finite total order on $E_{\mathcal{H}}$. We use the following axioms over a linearization $(\mathcal{H}, \mathsf{lin})$:

LinRYW. $\mathsf{so} \subseteq \mathsf{lin}$.

LinRT. $\mathsf{rt} \subseteq \mathsf{lin}$.

LinRetVal. $\forall e \in E.\, \mathsf{rval}(e) = \mathsf{eval}(\mathsf{pred}(e, \mathsf{lin}), \mathsf{oper}(e))$, where $\mathsf{pred}(e, \mathsf{lin})$ is the sequence of operations of events preceding $e$ in $\mathsf{lin}$.

Then the set of linearizable histories is given as follows:

$$\begin{aligned}\mathsf{HistLIN}_2 = \{(E, \mathsf{so}, \mathsf{rt}) \mid (\exists \mathsf{lin}.\, ((E, \mathsf{so}, \mathsf{rt}), \mathsf{lin}) \models \textsc{LinRYW} \wedge \textsc{LinRT} \wedge \textsc{LinRetVal}) \wedge \\ E \subseteq \mathsf{EPush} \cap \mathsf{EPull}\}.\end{aligned}$$

To ease stating the correspondence between the two definitions, the above definition includes the same constraints on fences as the definition in our framework, even though fences are not used in the corresponding axioms.

▶ **Proposition 23.** $\mathsf{HistLIN}_1 = \mathsf{HistLIN}_2$.

The proposition easily follows from the fact that in GSC executions where all operations push and pull, the vis relation is total and $\mathsf{rt} \subseteq \mathsf{ar}$.

## C.3 Correspondence with OSC

We partition the set of operations into read-only and update operations: $\mathsf{Op} = \mathsf{OpReadOnly} \uplus \mathsf{OpUpdate}$. Read-only operations do not change the state of an object:

$$\forall \xi, op.\, \mathsf{eval}(\xi, op) = \mathsf{eval}(\xi|_{\mathsf{OpUpdate}}, op),$$

where $\xi|_{\mathsf{OpUpdate}}$ is the projection of the context $\xi$ on $\mathsf{OpUpdate}$. We let $\mathsf{EReadOnly} = \{e \mid \mathsf{oper}(e) \in \mathsf{OpReadOnly}\}$ and $\mathsf{EUpdate} = \{e \mid \mathsf{oper}(e) \in \mathsf{OpUpdate}\}$.

In our framework, we can define the set of histories allowed by Ordered Sequential Consistency (OSC) as follows:

$$\mathsf{HistOSC}_1 = \{\mathcal{H} \mid \mathcal{H} \in \mathsf{HistGSC} \wedge E_{\mathcal{H}} \subseteq \mathsf{EPush} \wedge E_{\mathcal{H}} \cap \mathsf{EUpdate} \subseteq \mathsf{EPull}\}.$$

Thus, OSC is defined by making all events include push fences and update events additionally include pull fences.

We now give a reformulation of the OSC definition from [21]. We use the following axioms over a linearization $((E, \mathsf{so}, \mathsf{rt}), \mathsf{lin})$.

$\textsc{OscRYW}$. $\mathsf{so} \subseteq \mathsf{lin}$.

$\textsc{OscRT}$. $\mathsf{rt} \cap (\mathsf{Event} \times \mathsf{EUpdate}) \subseteq \mathsf{lin}$.

$\textsc{OscRetVal}$. $\forall e \in E.\, \mathsf{rval}(e) = \mathsf{eval}(\mathsf{pred}(e, \mathsf{lin}), \mathsf{oper}(e))$, where $\mathsf{pred}(e, \mathsf{lin})$ is the sequence of operations of events preceding $e$ in $\mathsf{lin}$.

Then the set of histories allowed by OSC according to the original definition is:

$$\mathsf{HistOSC}_2 = \{(E, \mathsf{so}, \mathsf{rt}) \mid (\exists \mathsf{lin}.\, ((E, \mathsf{so}, \mathsf{rt}), \mathsf{lin}) \models \textsc{OscRYW} \wedge \textsc{OscRT} \wedge \textsc{OscRetVal}) \wedge (E \subseteq \mathsf{EPush} \wedge E \cap \mathsf{EUpdate} \subseteq \mathsf{EPull})\}.$$

To ease stating the correspondence between the two definitions, the above definition includes the same constraints on fences as the definition in our framework, even though fences are not used in the corresponding axioms.

▶ **Proposition 24.** $\mathsf{HistOSC}_1 = \mathsf{HistOSC}_2$.

**Proof.** We first prove $\mathsf{HistOSC}_1 \subseteq \mathsf{HistOSC}_2$. Let $\mathcal{H} = (E, \mathsf{so}, \mathsf{rt}) \in \mathsf{HistOSC}_1$. Then

$$E \subseteq \mathsf{EPush} \wedge E \cap \mathsf{EUpdate} \subseteq \mathsf{EPull}$$

and for some $\mathsf{vis}$ and $\mathsf{ar}$ we have $(\mathcal{H}, \mathsf{vis}, \mathsf{ar}) \in \mathsf{ExecGSC}$. By $\textsc{PushedVis}$, we have

$$\mathsf{ar} \cap (\mathsf{EUpdate} \times \mathsf{EUpdate}) = \mathsf{vis} \cap (\mathsf{EUpdate} \times \mathsf{EUpdate}).$$

We let $\mathsf{lin}$ be the total order obtained by inserting events $e \in E \cap \mathsf{EReadOnly}$ into the above relation according to the following rule: an event $e$ goes after the last update event $f$ in $\mathsf{ar}$ such that $(f, e) \in \mathsf{vis}$, all events $e$ with the same corresponding event $f$ go in an arbitrary order consistent with $\mathsf{so}$. Then $\textsc{OscRetVal}$ follows from $\textsc{RetVal}$. Also, due to $\textsc{Eventual}$ and the prefix-finiteness of $\mathsf{ar}$, the relation $\mathsf{lin}$ is also prefix-finite.

We show $\textsc{OscRYW}$. Assume $e \xrightarrow{\mathsf{so}} f$; then by RYW we have $e \xrightarrow{\mathsf{vis}} f$. If $e, f \in \mathsf{EUpdate}$, then $e \xrightarrow{\mathsf{lin}} f$ follows from $e \xrightarrow{\mathsf{ar}} f$. If $e \in \mathsf{EUpdate}$ and $f \in \mathsf{EReadOnly}$, then $e \xrightarrow{\mathsf{lin}} f$ follows from $e \xrightarrow{\mathsf{vis}} f$. If $e \in \mathsf{EReadOnly}$ and $f \in \mathsf{EUpdate}$, then $e \xrightarrow{\mathsf{lin}} f$ follows from $\neg(f \xrightarrow{\mathsf{vis}} e)$, so that. Finally, consider the case when $e, f \in \mathsf{EReadOnly}$. If $e$ and $f$ see the same set of updates, then $e \xrightarrow{\mathsf{lin}} f$ by construction. Otherwise by $\textsc{MonotonicView}$ $f$ must see more updates than $e$. But then $f$ follows $e$ in $\mathsf{lin}$.

## APPENDIX

We show OscRT. Assume $e \xrightarrow{\mathsf{rt}} f$ and $f \in \mathsf{EUpdate}$. By PushedAr we get $e \xrightarrow{\mathsf{ar}} f$. Hence, if $e \in \mathsf{EUpdate}$, then $e \xrightarrow{\mathsf{lin}} f$. If $e \in \mathsf{EReadOnly}$, then by ObservedAr we cannot have $f \xrightarrow{\mathsf{vis}} e$. Hence, $e \xrightarrow{\mathsf{lin}} f$.

We now prove $\mathsf{HistOSC}_2 \subseteq \mathsf{HistOSC}_1$. Let $\mathcal{H} = (E, \mathsf{so}, \mathsf{rt}) \in \mathsf{HistOSC}_2$. Then for some total order $\mathsf{lin}$ on $E$, the axioms OscRYW, OscRT and OscRetVal hold and

$$E \subseteq \mathsf{EPush} \wedge E \cap \mathsf{EUpdate} \subseteq \mathsf{EPull}.$$

Let

$$R = \mathsf{lin} \cap (\mathsf{EUpdate} \times \mathsf{Event}).$$

By OscRT, we have $\mathsf{rt} \cap R^{-1} = \emptyset$. Then by Proposition 10 we have $\mathsf{rt}\ ;\ R\ ;\ \mathsf{rt} \subseteq \mathsf{rt}$. Hence, $R \cup \mathsf{rt}$ is acyclic and prefix-finite. Let $\mathsf{ar}$ be any prefix-finite total order containing $R \cup \mathsf{rt}$; such an order exists by Lemma 20. Let

$$\mathsf{vis} = \mathsf{so} \cup (\mathsf{ar}?\ ;\ (R \setminus \mathsf{so})\ ;\ \mathsf{so}?) \cup ((\mathsf{ar} \cap (\mathsf{Event} \times \mathsf{EUpdate}))\ ;\ \mathsf{so}?).$$

Then $\mathsf{vis} \subseteq \mathsf{ar}$ and, hence, $\mathsf{vis}$ is prefix-finite. Then $\mathcal{A} = (\mathcal{H}, \mathsf{vis}, \mathsf{ar})$ is an abstract execution. It is easy to check that $\mathcal{A}$ satisfies RYW, MonotonicView, ObservedVis, PushedVis, ObservedAr and PushedAr.

We now show that $\mathcal{A}$ satisfies RetVal. Consider $e \in E$. We show that $\mathsf{pred}(e, \mathsf{lin})$ and $\mathsf{ctxt}_{\mathcal{A}}(e)$ contain the same update events. Then by the definition of $\mathsf{ar}$ they contain them in the same order, which implies that

$$\mathsf{eval}(\mathsf{ctxt}_{\mathcal{A}}(e), \mathsf{oper}(e)) = \mathsf{eval}(\mathsf{pred}(e, \mathsf{lin}), \mathsf{oper}(e)).$$

Then RetVal follows from OscRetVal. We have $\mathsf{lin} \cap (\mathsf{EUpdate} \times \mathsf{Event}) = R \subseteq \mathsf{vis}$. Hence, any update event in $\mathsf{pred}(e, \mathsf{lin})$ is also in $\mathsf{ctxt}_{\mathcal{A}}(e)$. Consider now $f \in \mathsf{EUpdate}$ such that $f \xrightarrow{\mathsf{vis}} e$. Then

$$(f, e) \in \mathsf{so} \cup (\mathsf{ar}?\ ;\ (R \setminus \mathsf{so})\ ;\ \mathsf{so}?) \cup (\mathsf{ar} \cap (\mathsf{Event} \times \mathsf{EUpdate}))\ ;\ \mathsf{so}?.$$

If $(f, e) \in \mathsf{so}$, then by OscRYW we must have $(f, e) \in \mathsf{lin}$.

If $(f, e) \in \mathsf{ar}?\ ;\ (R \setminus \mathsf{so})\ ;\ \mathsf{so}?$, then for some $g, g'$ we have

$$f \xrightarrow{\mathsf{ar}?} g \xrightarrow{R \setminus \mathsf{so}} g' \xrightarrow{\mathsf{so}?} e.$$

Since $f \in \mathsf{EUpdate}$, this implies

$$f \xrightarrow{\mathsf{lin}?} g \xrightarrow{\mathsf{lin}} g' \xrightarrow{\mathsf{lin}?} e,$$

so that $(f, e) \in \mathsf{lin}$.

If $(f, e) \in (\mathsf{ar} \cap (\mathsf{Event} \times \mathsf{EUpdate}))\ ;\ \mathsf{so}?$, then for some $g \in \mathsf{EUpdate}$ we have

$$f \xrightarrow{\mathsf{ar}} g \xrightarrow{\mathsf{so}?} e.$$

Then we must have $(f, g) \in R$, so that $(f, g) \in \mathsf{lin}$ and $(f, e) \in \mathsf{lin}$.

Thus, in all cases we have $(f, e) \in \mathsf{lin}$, establishing RetVal.

Finally, we show that $\mathcal{A}$ satisfies Eventual. For any $e \in \mathsf{EUpdate}$, we have

$$|\{f \in E \mid \neg(e \xrightarrow{\mathsf{vis}} f)\}| < \infty, \tag{11}$$

because $R \subseteq \mathsf{vis}$. Assume that EVENTUAL is violated. Then for some $e \in \mathsf{EReadOnly}$, (11) does not hold. Let $e$ be the first such event in $\mathsf{ar}$. Assume that for some $g \in \mathsf{EUpdate}$ we have $e \xrightarrow{\mathsf{ar}} g$. Then for any $f \in E$ such that $g \xrightarrow{\mathsf{so}} f$ we have $e \xrightarrow{\mathsf{vis}} f$. Also, for any $f \in E$ such that $\neg(g \xrightarrow{\mathsf{so}} f)$ and $g \xrightarrow{\mathsf{lin}} f$ we have $e \xrightarrow{\mathsf{vis}} f$. Hence, (11) holds. Thus, we only need to consider the case when

$$\forall g.\, e \xrightarrow{\mathsf{ar}} g \implies g \in \mathsf{EReadOnly}.$$

By the definition of $\mathsf{vis}$, we have

$$\forall f, g.\, e \xrightarrow{\mathsf{ar}?} f \xrightarrow{\mathsf{ar}} g \implies (f \xrightarrow{\mathsf{vis}} g \iff f \xrightarrow{\mathsf{so}} g).$$

Let $e'$ be an event in $\mathsf{ar}$ such that $e \xrightarrow{\mathsf{ar}} e'$ and

$$\forall f.\, f \xrightarrow{\mathsf{ar}} e \implies f \xrightarrow{\mathsf{vis}} e'$$

Such an event $e'$ exists because we assume that $e$ is the first event in $\mathsf{ar}$ for which (11) does not hold. Let $\mathsf{vis}_1 = \mathsf{vis} \cup \{(e', g) \mid e' \xrightarrow{\mathsf{ar}?} g\}$. It is easy to check that the execution $(\mathcal{H}, \mathsf{vis}_1, \mathsf{ar})$ satisfies all axioms except possibly EVENTUAL, but where $e$ is now invisible to at most finitely many events. Continuing the above process of adding visibility edges *ad infinitum*, we can construct $\mathsf{vis}' \supseteq \mathsf{vis}$ such that $(\mathcal{H}, \mathsf{vis}', \mathsf{ar})$ satisfies all GSP axioms including EVENTUAL. It is also easy to check that $\mathsf{vis}'$ constructed in this way is prefix-finite, so $(\mathcal{H}, \mathsf{vis}', \mathsf{ar})$ is the desired execution. ◻

## D Proof of Theorem 2

Fix a well-fenced history $\mathcal{H} = (E, \mathsf{so}, \mathsf{rt})$ such that $\forall x.\, \mathcal{H}|_x \in \mathsf{HistGSC}$. Then for any $x$ there is an execution $\mathcal{A}_x = (\mathcal{H}|_x, \mathsf{vis}_x, \mathsf{ar}_x) \in \mathsf{ExecGSC}$, which we fix as well. Let

$$\mathsf{so}_0 = \bigcup_{x \in \mathsf{Obj}} \mathsf{so}_{\mathcal{H}|_x}; \quad \mathsf{rt}_0 = \bigcup_{x \in \mathsf{Obj}} \mathsf{rt}_{\mathcal{H}|_x}; \quad \mathsf{vis}_0 = \bigcup_{x \in \mathsf{Obj}} \mathsf{vis}_x; \quad \mathsf{ar}_0 = \bigcup_{x \in \mathsf{Obj}} \mathsf{ar}_x.$$

Then the tuple $((E, \mathsf{so}_0, \mathsf{rt}_0), \mathsf{vis}_0, \mathsf{ar}_0)$ satisfies all the GSC axioms except for possibly EVENTUAL (even though this tuple is not a well-formed execution). In the following, all mentions of axioms refer to the above tuple unless otherwise specified.

For a set $E_0 \subseteq E$ let $\langle E_0 \rangle = \mathsf{Id} \cap (E_0 \times E_0)$. Then we can rewrite the definition of $\prec$ in the following more concise form:

$$\begin{aligned} e \prec f \iff & \exists g.\, \mathsf{obj}(f) = \mathsf{obj}(g) \wedge (f, g) \notin \mathsf{vis}_0 \wedge \\ & (e, g) \in ((\mathsf{vis}_0 \setminus \mathsf{so}) \cup \langle \mathsf{EPush} \rangle) \,;\, (\mathsf{rt} \cap (\mathsf{Event} \times \mathsf{EPull})) \,;\, \mathsf{so}_0?. \end{aligned}$$

We start by proving Lemma 7, which gives a closed-form expression covering $R^+$. We build the closed form gradually, by considering subsets of $R$.

▶ **Proposition 25.** $\mathsf{so} \subseteq \mathsf{ar}_0 \cup \mathsf{ar}_0? \,;\, \overline{\mathsf{rt}}$.

**Proof.** Assume $e \xrightarrow{\mathsf{so}} f$. If $\mathsf{obj}(e) = \mathsf{obj}(f)$, then by RYW we have $e \xrightarrow{\mathsf{ar}_0} f$. Otherwise, since the history $\mathcal{H}$ is well-fenced, for some $e' \in \mathsf{EPush}$ such that $\mathsf{obj}(e') = \mathsf{obj}(e)$ we have $e \xrightarrow{\mathsf{so}?} e' \xrightarrow{\mathsf{so}} f$. Then $e \xrightarrow{\mathsf{ar}_0?} e' \xrightarrow{\overline{\mathsf{rt}}} f$, as required. ◻

▶ **Proposition 26.** $(\overline{\mathsf{rt}} \cup \mathsf{so})^+ \subseteq \overline{\mathsf{rt}} \cup \mathsf{so} \,;\, \overline{\mathsf{rt}}?$.

**Proof.** Follows from the fact that $\overline{\mathsf{rt}} \,;\, \mathsf{so} \subseteq \overline{\mathsf{rt}}$. ◻

## APPENDIX

▶ **Proposition 27.**

$$(\overline{\mathsf{rt}} \cup \mathsf{so} \cup \mathsf{ar}_0)^+ \subseteq \overline{\mathsf{rt}} \cup \mathsf{ar}_0 \cup \mathsf{ar}_0 ; \overline{\mathsf{rt}} \cup \overline{\mathsf{rt}} ; \mathsf{ar}_0 \cup \mathsf{ar}_0 ; \overline{\mathsf{rt}} ; \mathsf{ar}_0.$$

**Proof**. We have:

$$\begin{aligned}
\mathsf{ar}_0 ; (\overline{\mathsf{rt}} \cup \mathsf{so})^+ &\subseteq \mathsf{ar}_0 ; (\overline{\mathsf{rt}} \cup \mathsf{so} ; \overline{\mathsf{rt}}?) && \text{by Proposition 26} \\
&= \mathsf{ar}_0 ; \overline{\mathsf{rt}} \cup \mathsf{ar}_0 ; \mathsf{so} ; \overline{\mathsf{rt}}? \\
&\subseteq \mathsf{ar}_0 ; \overline{\mathsf{rt}} \cup \mathsf{ar}_0 ; (\mathsf{ar}_0 \cup \mathsf{ar}_0? ; \overline{\mathsf{rt}}) ; \overline{\mathsf{rt}}? && \text{by Proposition 25} \\
&= \mathsf{ar}_0 ; \overline{\mathsf{rt}}?
\end{aligned}$$

Thus,

$$\mathsf{ar}_0 ; (\overline{\mathsf{rt}} \cup \mathsf{so})^+ \subseteq \mathsf{ar}_0 ; \overline{\mathsf{rt}}?. \tag{12}$$

Then

$$\begin{aligned}
& (\overline{\mathsf{rt}} \cup \mathsf{so} \cup \mathsf{ar}_0)^+ \\
= {} & (\overline{\mathsf{rt}} \cup \mathsf{so})^+ \cup (\overline{\mathsf{rt}} \cup \mathsf{so})^* ; (\mathsf{ar}_0 ; (\overline{\mathsf{rt}} \cup \mathsf{so})^*)^+ \\
\subseteq {} & (\overline{\mathsf{rt}} \cup \mathsf{so} ; \overline{\mathsf{rt}}?) \cup (\overline{\mathsf{rt}} \cup \mathsf{so} ; \overline{\mathsf{rt}}?)? ; (\mathsf{ar}_0 ; \overline{\mathsf{rt}}?)^+ && \text{by Proposition 26 and (12)} \\
\subseteq {} & \overline{\mathsf{rt}} \cup \mathsf{so} ; \overline{\mathsf{rt}}? \cup (\overline{\mathsf{rt}} \cup \mathsf{so} ; \overline{\mathsf{rt}}?)? ; (\mathsf{ar}_0 ; \overline{\mathsf{rt}}? \cup \mathsf{ar}_0 ; \overline{\mathsf{rt}} ; \mathsf{ar}_0) && \text{by Corollary 11} \\
= {} & \overline{\mathsf{rt}} \cup \mathsf{so} ; \overline{\mathsf{rt}}? \cup \mathsf{ar}_0 ; \overline{\mathsf{rt}}? \cup \mathsf{ar}_0 ; \overline{\mathsf{rt}} ; \mathsf{ar}_0 \cup \\
& (\overline{\mathsf{rt}} \cup \mathsf{so} ; \overline{\mathsf{rt}}?) ; (\mathsf{ar}_0 ; \overline{\mathsf{rt}}? \cup \mathsf{ar}_0 ; \overline{\mathsf{rt}} ; \mathsf{ar}_0) \\
\subseteq {} & \overline{\mathsf{rt}} \cup \mathsf{so} ; \overline{\mathsf{rt}}? \cup \mathsf{ar}_0 ; \overline{\mathsf{rt}}? \cup \mathsf{ar}_0 ; \overline{\mathsf{rt}} ; \mathsf{ar}_0 \cup \\
& \overline{\mathsf{rt}} ; \mathsf{ar}_0 \cup \mathsf{so} ; \mathsf{ar}_0 ; \overline{\mathsf{rt}}? \cup \mathsf{so} ; \overline{\mathsf{rt}} ; \mathsf{ar}_0 \cup \mathsf{so} ; \mathsf{ar}_0 ; \overline{\mathsf{rt}} ; \mathsf{ar}_0 && \text{by Corollary 11} \\
\subseteq {} & \overline{\mathsf{rt}} \cup (\mathsf{ar}_0 \cup \mathsf{ar}_0? ; \overline{\mathsf{rt}}) ; \overline{\mathsf{rt}}? \cup \mathsf{ar}_0 ; \overline{\mathsf{rt}}? \cup \mathsf{ar}_0 ; \overline{\mathsf{rt}} ; \mathsf{ar}_0 \cup \\
& \overline{\mathsf{rt}} ; \mathsf{ar}_0 \cup (\mathsf{ar}_0 \cup \mathsf{ar}_0? ; \overline{\mathsf{rt}}) ; \mathsf{ar}_0 ; \overline{\mathsf{rt}}? \cup \\
& (\mathsf{ar}_0 \cup \mathsf{ar}_0? ; \overline{\mathsf{rt}}) ; \overline{\mathsf{rt}} ; \mathsf{ar}_0 \cup \\
& (\mathsf{ar}_0 \cup \mathsf{ar}_0? ; \overline{\mathsf{rt}}) ; \mathsf{ar}_0 ; \overline{\mathsf{rt}} ; \mathsf{ar}_0 && \text{by Proposition 25} \\
\subseteq {} & \overline{\mathsf{rt}} \cup \mathsf{ar}_0 \cup \mathsf{ar}_0 ; \overline{\mathsf{rt}} \cup \overline{\mathsf{rt}} ; \mathsf{ar}_0 \cup \mathsf{ar}_0 ; \overline{\mathsf{rt}} ; \mathsf{ar}_0 && \text{by Corollary 11}
\end{aligned}$$

□

▶ **Proposition 28.**

$$\begin{aligned}
& (\overline{\mathsf{rt}} \cup \mathsf{so} \cup \mathsf{ar}_0 \cup ((\mathsf{vis}_0 \setminus \mathsf{so}) ; \mathsf{rt}))^+ \\
\subseteq {} & \overline{\mathsf{rt}} \cup \mathsf{ar}_0 \cup \mathsf{ar}_0 ; \overline{\mathsf{rt}} \cup \overline{\mathsf{rt}} ; \mathsf{ar}_0 \cup \mathsf{ar}_0 ; \overline{\mathsf{rt}} ; \mathsf{ar}_0 \cup \\
& (\mathsf{vis}_0 \setminus \mathsf{so}) ; \mathsf{rt} \cup \mathsf{ar}_0 ; (\mathsf{vis}_0 \setminus \mathsf{so}) ; \mathsf{rt} \cup (\mathsf{vis}_0 \setminus \mathsf{so}) ; \mathsf{rt} ; \mathsf{ar}_0 \cup \mathsf{ar}_0 ; (\mathsf{vis}_0 \setminus \mathsf{so}) ; \mathsf{rt} ; \mathsf{ar}_0.
\end{aligned}$$

**Proof**. Applying first Proposition 27 and then Corollary 11 and ObservedVis, we get:

$$(\overline{\mathsf{rt}} \cup \mathsf{so} \cup \mathsf{ar}_0)^* ; ((\mathsf{vis}_0 \setminus \mathsf{so}) ; \mathsf{rt}) \subseteq$$

$$\begin{array}{lclcl}
(\mathsf{vis}_0 \setminus \mathsf{so}) ; \mathsf{rt} \cup & \subseteq & (\mathsf{vis}_0 \setminus \mathsf{so}) ; \mathsf{rt} \cup & \subseteq & (\mathsf{vis}_0 \setminus \mathsf{so}) ; \mathsf{rt} \cup \\
\overline{\mathsf{rt}} ; (\mathsf{vis}_0 \setminus \mathsf{so}) ; \mathsf{rt} \cup & & \overline{\mathsf{rt}} \cup & & \overline{\mathsf{rt}} \cup \\
\mathsf{ar}_0 ; (\mathsf{vis}_0 \setminus \mathsf{so}) ; \mathsf{rt} \cup & & \mathsf{ar}_0 ; (\mathsf{vis}_0 \setminus \mathsf{so}) ; \mathsf{rt} \cup & & \mathsf{ar}_0 ; (\mathsf{vis}_0 \setminus \mathsf{so}) ; \mathsf{rt} \cup \\
\mathsf{ar}_0 ; \overline{\mathsf{rt}} ; (\mathsf{vis}_0 \setminus \mathsf{so}) ; \mathsf{rt} \cup & & \mathsf{ar}_0 ; \overline{\mathsf{rt}} \cup & & \mathsf{ar}_0 ; \overline{\mathsf{rt}} \\
\overline{\mathsf{rt}} ; \mathsf{ar}_0 ; (\mathsf{vis}_0 \setminus \mathsf{so}) ; \mathsf{rt} \cup & & \overline{\mathsf{rt}} \cup & & \\
\mathsf{ar}_0 ; \overline{\mathsf{rt}} ; \mathsf{ar}_0 ; (\mathsf{vis}_0 \setminus \mathsf{so}) ; \mathsf{rt} & & \mathsf{ar}_0 ; \overline{\mathsf{rt}} & &
\end{array}$$

Thus,

$$(\overline{\mathsf{rt}} \cup \mathsf{so} \cup \mathsf{ar}_0)^* ; ((\mathsf{vis}_0 \setminus \mathsf{so}) ; \mathsf{rt}) \subseteq (\mathsf{vis}_0 \setminus \mathsf{so}) ; \mathsf{rt} \cup \overline{\mathsf{rt}} \cup \mathsf{ar}_0 ; (\mathsf{vis}_0 \setminus \mathsf{so}) ; \mathsf{rt} \cup \mathsf{ar}_0 ; \overline{\mathsf{rt}}. \tag{13}$$

The relation on the right-hand side of the above inclusion is transitive:

$$
\begin{aligned}
& ((\mathsf{vis}_0 \setminus \mathsf{so}) \mathbin{;} \mathsf{rt} \cup \overline{\mathsf{rt}} \cup \mathsf{ar}_0 \mathbin{;} (\mathsf{vis}_0 \setminus \mathsf{so}) \mathbin{;} \mathsf{rt} \cup \mathsf{ar}_0 \mathbin{;} \overline{\mathsf{rt}}) \mathbin{;} \\
& ((\mathsf{vis}_0 \setminus \mathsf{so}) \mathbin{;} \mathsf{rt} \cup \overline{\mathsf{rt}} \cup \mathsf{ar}_0 \mathbin{;} (\mathsf{vis}_0 \setminus \mathsf{so}) \mathbin{;} \mathsf{rt} \cup \mathsf{ar}_0 \mathbin{;} \overline{\mathsf{rt}}) \\
\subseteq\; & ((\mathsf{vis}_0 \setminus \mathsf{so}) \mathbin{;} \mathsf{rt} \cup \overline{\mathsf{rt}} \cup \mathsf{ar}_0 \mathbin{;} (\mathsf{vis}_0 \setminus \mathsf{so}) \mathbin{;} \mathsf{rt} \cup \mathsf{ar}_0 \mathbin{;} \overline{\mathsf{rt}}) \mathbin{;} && \text{by } \textsc{ObservedVis} \\
& ((\mathsf{vis}_0 \setminus \mathsf{so}) \mathbin{;} \mathsf{rt} \cup \overline{\mathsf{rt}} \cup \mathsf{vis}_0 \mathbin{;} \mathsf{rt} \cup \mathsf{ar}_0 \mathbin{;} \overline{\mathsf{rt}}) \\
\subseteq\; & (\mathsf{vis}_0 \setminus \mathsf{so}) \mathbin{;} \mathsf{rt} \cup \overline{\mathsf{rt}} \cup \mathsf{ar}_0 \mathbin{;} (\mathsf{vis}_0 \setminus \mathsf{so}) \mathbin{;} \mathsf{rt} \cup \mathsf{ar}_0 \mathbin{;} \overline{\mathsf{rt}} && \text{by Corollary 11}
\end{aligned}
$$

Using this fact and (13), we get

$$
((\overline{\mathsf{rt}} \cup \mathsf{so} \cup \mathsf{ar}_0)^* \mathbin{;} ((\mathsf{vis}_0 \setminus \mathsf{so}) \mathbin{;} \mathsf{rt}))^+ \subseteq (\mathsf{vis}_0 \setminus \mathsf{so}) \mathbin{;} \mathsf{rt} \cup \overline{\mathsf{rt}} \cup \mathsf{ar}_0 \mathbin{;} (\mathsf{vis}_0 \setminus \mathsf{so}) \mathbin{;} \mathsf{rt} \cup \mathsf{ar}_0 \mathbin{;} \overline{\mathsf{rt}}. \quad (14)
$$

Then

$$
\begin{aligned}
& (\overline{\mathsf{rt}} \cup \mathsf{so} \cup \mathsf{ar}_0 \cup ((\mathsf{vis}_0 \setminus \mathsf{so}) \mathbin{;} \mathsf{rt}))^+ \\
=\; & (\overline{\mathsf{rt}} \cup \mathsf{so} \cup \mathsf{ar}_0)^+ \cup ((\overline{\mathsf{rt}} \cup \mathsf{so} \cup \mathsf{ar}_0)^* \mathbin{;} ((\mathsf{vis}_0 \setminus \mathsf{so}) \mathbin{;} \mathsf{rt}))^+ \mathbin{;} (\overline{\mathsf{rt}} \cup \mathsf{so} \cup \mathsf{ar}_0)^* \\
\subseteq\; & (\overline{\mathsf{rt}} \cup \mathsf{ar}_0 \cup \mathsf{ar}_0 \mathbin{;} \overline{\mathsf{rt}} \cup \overline{\mathsf{rt}} \mathbin{;} \mathsf{ar}_0 \cup \mathsf{ar}_0 \mathbin{;} \overline{\mathsf{rt}} \mathbin{;} \mathsf{ar}_0) \cup && \text{by (14)} \\
& ((\mathsf{vis}_0 \setminus \mathsf{so}) \mathbin{;} \mathsf{rt} \cup \overline{\mathsf{rt}} \cup \mathsf{ar}_0 \mathbin{;} (\mathsf{vis}_0 \setminus \mathsf{so}) \mathbin{;} \mathsf{rt} \cup \mathsf{ar}_0 \mathbin{;} \overline{\mathsf{rt}}) \mathbin{;} \\
& (\overline{\mathsf{rt}} \cup \mathsf{ar}_0 \cup \mathsf{ar}_0 \mathbin{;} \overline{\mathsf{rt}} \cup \overline{\mathsf{rt}} \mathbin{;} \mathsf{ar}_0 \cup \mathsf{ar}_0 \mathbin{;} \overline{\mathsf{rt}} \mathbin{;} \mathsf{ar}_0)? \\
=\; & \overline{\mathsf{rt}} \cup \mathsf{ar}_0 \cup \mathsf{ar}_0 \mathbin{;} \overline{\mathsf{rt}} \cup \overline{\mathsf{rt}} \mathbin{;} \mathsf{ar}_0 \cup \mathsf{ar}_0 \mathbin{;} \overline{\mathsf{rt}} \mathbin{;} \mathsf{ar}_0 \cup \\
& (\mathsf{vis}_0 \setminus \mathsf{so}) \mathbin{;} \mathsf{rt} \cup \overline{\mathsf{rt}} \cup \mathsf{ar}_0 \mathbin{;} (\mathsf{vis}_0 \setminus \mathsf{so}) \mathbin{;} \mathsf{rt} \cup \mathsf{ar}_0 \mathbin{;} \overline{\mathsf{rt}} \cup \\
& ((\mathsf{vis}_0 \setminus \mathsf{so}) \mathbin{;} \mathsf{rt} \cup \overline{\mathsf{rt}} \cup \mathsf{ar}_0 \mathbin{;} (\mathsf{vis}_0 \setminus \mathsf{so}) \mathbin{;} \mathsf{rt} \cup \mathsf{ar}_0 \mathbin{;} \overline{\mathsf{rt}}) \mathbin{;} \\
& (\overline{\mathsf{rt}} \cup \mathsf{ar}_0 \cup \mathsf{ar}_0 \mathbin{;} \overline{\mathsf{rt}} \cup \overline{\mathsf{rt}} \mathbin{;} \mathsf{ar}_0 \cup \mathsf{ar}_0 \mathbin{;} \overline{\mathsf{rt}} \mathbin{;} \mathsf{ar}_0) \\
\subseteq\; & \overline{\mathsf{rt}} \cup \mathsf{ar}_0 \cup \mathsf{ar}_0 \mathbin{;} \overline{\mathsf{rt}} \cup \overline{\mathsf{rt}} \mathbin{;} \mathsf{ar}_0 \cup \mathsf{ar}_0 \mathbin{;} \overline{\mathsf{rt}} \mathbin{;} \mathsf{ar}_0 \cup && \text{by Corollary 11} \\
& (\mathsf{vis}_0 \setminus \mathsf{so}) \mathbin{;} \mathsf{rt} \cup \mathsf{ar}_0 \mathbin{;} (\mathsf{vis}_0 \setminus \mathsf{so}) \mathbin{;} \mathsf{rt} \cup \\
& ((\mathsf{vis}_0 \setminus \mathsf{so}) \mathbin{;} \mathsf{rt} \cup \overline{\mathsf{rt}} \cup \mathsf{ar}_0 \mathbin{;} (\mathsf{vis}_0 \setminus \mathsf{so}) \mathbin{;} \mathsf{rt} \cup \mathsf{ar}_0 \mathbin{;} \overline{\mathsf{rt}}) \mathbin{;} \mathsf{ar}_0 \\
=\; & \overline{\mathsf{rt}} \cup \mathsf{ar}_0 \cup \mathsf{ar}_0 \mathbin{;} \overline{\mathsf{rt}} \cup \overline{\mathsf{rt}} \mathbin{;} \mathsf{ar}_0 \cup \mathsf{ar}_0 \mathbin{;} \overline{\mathsf{rt}} \mathbin{;} \mathsf{ar}_0 \cup \\
& (\mathsf{vis}_0 \setminus \mathsf{so}) \mathbin{;} \mathsf{rt} \cup \mathsf{ar}_0 \mathbin{;} (\mathsf{vis}_0 \setminus \mathsf{so}) \mathbin{;} \mathsf{rt} \cup \\
& (\mathsf{vis}_0 \setminus \mathsf{so}) \mathbin{;} \mathsf{rt} \mathbin{;} \mathsf{ar}_0 \cup \overline{\mathsf{rt}} \mathbin{;} \mathsf{ar}_0 \cup \mathsf{ar}_0 \mathbin{;} (\mathsf{vis}_0 \setminus \mathsf{so}) \mathbin{;} \mathsf{rt} \mathbin{;} \mathsf{ar}_0 \cup \mathsf{ar}_0 \mathbin{;} \overline{\mathsf{rt}} \mathbin{;} \mathsf{ar}_0 \\
=\; & \overline{\mathsf{rt}} \cup \mathsf{ar}_0 \cup \mathsf{ar}_0 \mathbin{;} \overline{\mathsf{rt}} \cup \overline{\mathsf{rt}} \mathbin{;} \mathsf{ar}_0 \cup \mathsf{ar}_0 \mathbin{;} \overline{\mathsf{rt}} \mathbin{;} \mathsf{ar}_0 \cup \\
& (\mathsf{vis}_0 \setminus \mathsf{so}) \mathbin{;} \mathsf{rt} \cup \mathsf{ar}_0 \mathbin{;} (\mathsf{vis}_0 \setminus \mathsf{so}) \mathbin{;} \mathsf{rt} \cup \\
& (\mathsf{vis}_0 \setminus \mathsf{so}) \mathbin{;} \mathsf{rt} \mathbin{;} \mathsf{ar}_0 \cup \mathsf{ar}_0 \mathbin{;} (\mathsf{vis}_0 \setminus \mathsf{so}) \mathbin{;} \mathsf{rt} \mathbin{;} \mathsf{ar}_0
\end{aligned}
$$

◻

▶ **Lemma 29.** $\overline{\mathsf{rt}} \cup \mathsf{so} \cup \mathsf{ar}_0 \cup ((\mathsf{vis}_0 \setminus \mathsf{so}) \mathbin{;} \mathsf{rt})$ *is acyclic.*

**Proof.** This easily follows from Proposition 28 and the axioms ObservedAr and PushedAr. ◻

▶ **Proposition 30.**

- $\overline{\mathsf{rt}} \mathbin{;} \prec \subseteq \prec$.
- $\overline{\mathsf{rt}} \mathbin{;} \mathsf{ar}_0 \mathbin{;} \prec \subseteq \prec$.
- $(\mathsf{vis}_0 \setminus \mathsf{so}) \mathbin{;} \mathsf{rt} \mathbin{;} \prec \subseteq \prec$.
- $(\mathsf{vis}_0 \setminus \mathsf{so}) \mathbin{;} \mathsf{rt} \mathbin{;} \mathsf{ar}_0 \mathbin{;} \prec \subseteq \prec$.

**Proof.** We only prove the second property; the others are proved analogously. Assume $e \xrightarrow{\overline{\mathsf{rt}};\mathsf{ar}_0} e' \prec f'$. Then for some $g$ we have $\mathsf{obj}(f') = \mathsf{obj}(g)$ and

$$
(e', g) \in ((\mathsf{vis}_0 \setminus \mathsf{so}) \cup \langle \mathsf{EPush} \rangle) \mathbin{;} (\mathsf{rt} \cap (\mathsf{Event} \times \mathsf{EPull})) \mathbin{;} \mathsf{so}_0? \wedge (f', g) \notin \mathsf{vis}_0.
$$

Then

$$
(e, g) \in \overline{\mathsf{rt}} \mathbin{;} \mathsf{ar}_0 \mathbin{;} ((\mathsf{vis}_0 \setminus \mathsf{so}) \cup \langle \mathsf{EPush} \rangle) \mathbin{;} (\mathsf{rt} \cap (\mathsf{Event} \times \mathsf{EPull})) \mathbin{;} \mathsf{so}_0?.
$$

By ObservedVis this entails

$$(e, g) \in \overline{\mathsf{rt}} \ ; (\mathsf{vis}_0 \cup \overline{\mathsf{ar}}_0) \ ; (\mathsf{rt} \cap (\mathsf{Event} \times \mathsf{EPull})) \ ; \mathsf{so}_0?.$$

Then by Corollary 11 we get

$$(e, g) \in \langle \mathsf{EPush} \rangle \ ; (\mathsf{rt} \cap (\mathsf{Event} \times \mathsf{EPull})) \ ; \mathsf{so}_0?,$$

which together with $(f', g) \notin \mathsf{vis}_0$ implies $e \prec f'$. ◻

▶ **Proposition 31.** $\prec \ ; \mathsf{ar}_0? \ ; \prec \ \subseteq \ \prec$.

**Proof**. Assume

$$e_1 \prec f_1 \xrightarrow{\mathsf{ar}_0?} e_2 \prec f_2.$$

Then for some $g_1, g_2, e'_1, e'_2, g'_1, g'_2$ we have $\mathsf{obj}(g_1) = \mathsf{obj}(f_1)$, $\mathsf{obj}(g_2) = \mathsf{obj}(f_2)$ and

$$\neg(f_1 \xrightarrow{\mathsf{vis}_0} g_1) \wedge e_1 \xrightarrow{(\mathsf{vis}_0 \setminus \mathsf{so}) \cup \langle \mathsf{EPush} \rangle} e'_1 \xrightarrow{\mathsf{rt} \cap (\mathsf{Event} \times \mathsf{EPull})} g'_1 \xrightarrow{\mathsf{so}_0?} g_1;$$
$$\neg(f_2 \xrightarrow{\mathsf{vis}_0} g_2) \wedge e_2 \xrightarrow{(\mathsf{vis}_0 \setminus \mathsf{so}) \cup \langle \mathsf{EPush} \rangle} e'_2 \xrightarrow{\mathsf{rt} \cap (\mathsf{Event} \times \mathsf{EPull})} g'_2 \xrightarrow{\mathsf{so}_0?} g_2.$$

Since $\mathsf{rt}$ is an interval order, either $e'_1 \xrightarrow{\mathsf{rt}} g'_2$ or $e'_2 \xrightarrow{\mathsf{rt}} g'_1$. If $e'_2 \xrightarrow{\mathsf{rt}} g'_1$, then

$$f_1 \xrightarrow{\mathsf{ar}_0?} e_2 \xrightarrow{(\mathsf{vis}_0 \setminus \mathsf{so}) \cup \langle \mathsf{EPush} \rangle} e'_2 \xrightarrow{\mathsf{rt} \cap (\mathsf{Event} \times \mathsf{EPull})} g'_1 \xrightarrow{\mathsf{so}_0?} g_1.$$

We have

$$\mathsf{obj}(g'_1) = \mathsf{obj}(g_1) = \mathsf{obj}(f_1) = \mathsf{obj}(e_2) = \mathsf{obj}(e'_2).$$

Then by ObservedVis, PushedVis and MonotonicView we get $e_2 \xrightarrow{\mathsf{vis}_0} g_1$, yielding a contradiction. Hence, we cannot have $e'_2 \xrightarrow{\mathsf{rt}} g'_1$, so we must have $e'_1 \xrightarrow{\mathsf{rt}} g'_2$. Then

$$\neg(f_2 \xrightarrow{\mathsf{vis}_0} g_2) \wedge e_1 \xrightarrow{(\mathsf{vis}_0 \setminus \mathsf{so}) \cup \langle \mathsf{EPush} \rangle} e'_1 \xrightarrow{\mathsf{rt} \cap (\mathsf{Event} \times \mathsf{EPull})} g'_2 \xrightarrow{\mathsf{so}_0?} g_2,$$

so that $e_1 \prec f_2$. ◻

▶ **Proposition 32.**

$$\begin{aligned} & (\overline{\mathsf{rt}} \cup \mathsf{so} \cup \mathsf{ar}_0 \cup ((\mathsf{vis}_0 \setminus \mathsf{so}) \ ; \mathsf{rt}) \cup \prec)^+ \\ = \ & (\overline{\mathsf{rt}} \cup \mathsf{so} \cup \mathsf{ar}_0 \cup ((\mathsf{vis}_0 \setminus \mathsf{so}) \ ; \mathsf{rt}))^+ \cup (\prec \cup \, \mathsf{ar}_0 \ ; \prec) \ ; (\overline{\mathsf{rt}} \cup \mathsf{so} \cup \mathsf{ar}_0 \cup ((\mathsf{vis}_0 \setminus \mathsf{so}) \ ; \mathsf{rt}))^*. \end{aligned}$$

**Proof**. We have

$$\begin{aligned} & (\overline{\mathsf{rt}} \cup \mathsf{so} \cup \mathsf{ar}_0 \cup ((\mathsf{vis}_0 \setminus \mathsf{so}) \ ; \mathsf{rt}))^+ \ ; \prec \\ \subseteq \ & (\overline{\mathsf{rt}} \cup \mathsf{ar}_0 \cup \mathsf{ar}_0 \ ; \overline{\mathsf{rt}} \cup \overline{\mathsf{rt}} \ ; \mathsf{ar}_0 \cup \mathsf{ar}_0 \ ; \overline{\mathsf{rt}} \ ; \mathsf{ar}_0 \cup (\mathsf{vis}_0 \setminus \mathsf{so}) \ ; \mathsf{rt} \ \cup && \text{by Proposition 28} \\ & \mathsf{ar}_0 \ ; (\mathsf{vis}_0 \setminus \mathsf{so}) \ ; \mathsf{rt} \cup (\mathsf{vis}_0 \setminus \mathsf{so}) \ ; \mathsf{rt} \ ; \mathsf{ar}_0 \ \cup \\ & \mathsf{ar}_0 \ ; (\mathsf{vis}_0 \setminus \mathsf{so}) \ ; \mathsf{rt} \ ; \mathsf{ar}_0) \ ; \prec \\ \subseteq \ & \prec \cup \, \mathsf{ar}_0 \ ; \prec && \text{by Proposition 30} \end{aligned}$$

Thus,

$$(\overline{\mathsf{rt}} \cup \mathsf{so} \cup \mathsf{ar}_0 \cup ((\mathsf{vis}_0 \setminus \mathsf{so}) \ ; \mathsf{rt}))^+ \ ; \prec \ \subseteq \ \prec \cup \, \mathsf{ar}_0 \ ; \prec. \tag{15}$$

Then

$$\begin{aligned}
&(\overline{\mathsf{rt}} \cup \mathsf{so} \cup \mathsf{ar}_0 \cup ((\mathsf{vis}_0 \setminus \mathsf{so}) \; ; \; \mathsf{rt}) \cup \prec)^+ \\
= \; &(\overline{\mathsf{rt}} \cup \mathsf{so} \cup \mathsf{ar}_0 \cup ((\mathsf{vis}_0 \setminus \mathsf{so}) \; ; \; \mathsf{rt}))^+ \cup \\
&((\overline{\mathsf{rt}} \cup \mathsf{so} \cup \mathsf{ar}_0 \cup ((\mathsf{vis}_0 \setminus \mathsf{so}) \; ; \; \mathsf{rt}))^* \; ; \prec)^+ \; ; \\
&\quad (\overline{\mathsf{rt}} \cup \mathsf{so} \cup \mathsf{ar}_0 \cup ((\mathsf{vis}_0 \setminus \mathsf{so}) \; ; \; \mathsf{rt}))^* \\
\subseteq \; &(\overline{\mathsf{rt}} \cup \mathsf{so} \cup \mathsf{ar}_0 \cup ((\mathsf{vis}_0 \setminus \mathsf{so}) \; ; \; \mathsf{rt}))^+ \cup && \text{by (15)} \\
&(\prec \cup \, \mathsf{ar}_0 \; ; \prec)^+ \; ; (\overline{\mathsf{rt}} \cup \mathsf{so} \cup \mathsf{ar}_0 \cup ((\mathsf{vis}_0 \setminus \mathsf{so}) \; ; \; \mathsf{rt}))^* \\
\subseteq \; &(\overline{\mathsf{rt}} \cup \mathsf{so} \cup \mathsf{ar}_0 \cup ((\mathsf{vis}_0 \setminus \mathsf{so}) \; ; \; \mathsf{rt}))^+ \cup && \text{by Proposition 31} \\
&(\prec \cup \, \mathsf{ar}_0 \; ; \prec) \; ; (\overline{\mathsf{rt}} \cup \mathsf{so} \cup \mathsf{ar}_0 \cup ((\mathsf{vis}_0 \setminus \mathsf{so}) \; ; \; \mathsf{rt}))^*
\end{aligned}$$

◻

Lemma 7 follows from Proposition 28 and 32.

▶ **Proposition 33.** $\mathsf{ar}_0? \; ; \prec$ *is irreflexive.*

**Proof.** Assume that for some $e, e'$ we have $e \xrightarrow{\mathsf{ar}_0?} e' \prec e$. Hence, for some $g$ such that $\mathsf{obj}(g) = \mathsf{obj}(e)$ we have

$$(e, g) \notin \mathsf{vis}_0 \wedge (e', g) \in ((\mathsf{vis}_0 \setminus \mathsf{so}) \cup \langle \mathsf{EPush} \rangle) \; ; (\mathsf{rt} \cap (\mathsf{Event} \times \mathsf{EPull})) \; ; \mathsf{so}_0?.$$

Then

$$(e, g) \in (\mathsf{ar}_0? \; ; (\mathsf{vis}_0 \setminus \mathsf{so}) \cup \langle \mathsf{EPush} \rangle) \; ; (\mathsf{rt} \cap (\mathsf{Event} \times \mathsf{EPull})) \; ; \mathsf{so}_0?.$$

Since $\mathsf{obj}(g) = \mathsf{obj}(e)$, by ObservedVis, PushedVis and MonotonicView we get $(e, g) \in \mathsf{vis}_0$, yielding a contradiction. ◻

**Proof of Lemma 6.** Assume that for some $e$,

$$(e, e) \in (\overline{\mathsf{rt}} \cup \mathsf{so} \cup \mathsf{ar}_0 \cup ((\mathsf{vis}_0 \setminus \mathsf{so}) \; ; \; \mathsf{rt}) \cup \prec)^+.$$

Then by Proposition 32 we have

$$(e, e) \in (\overline{\mathsf{rt}} \cup \mathsf{so} \cup \mathsf{ar}_0 \cup ((\mathsf{vis}_0 \setminus \mathsf{so}) \; ; \; \mathsf{rt}))^+ \cup (\prec \cup \, \mathsf{ar}_0 \; ; \prec) \; ; (\overline{\mathsf{rt}} \cup \mathsf{so} \cup \mathsf{ar}_0 \cup ((\mathsf{vis}_0 \setminus \mathsf{so}) \; ; \; \mathsf{rt}))^*.$$

Hence, for some $f$,

$$\begin{aligned}
(f, f) \in \; &(\overline{\mathsf{rt}} \cup \mathsf{so} \cup \mathsf{ar}_0 \cup ((\mathsf{vis}_0 \setminus \mathsf{so}) \; ; \; \mathsf{rt}))^+ \cup \\
&(\overline{\mathsf{rt}} \cup \mathsf{so} \cup \mathsf{ar}_0 \cup ((\mathsf{vis}_0 \setminus \mathsf{so}) \; ; \; \mathsf{rt}))^* \; ; (\prec \cup \, \mathsf{ar}_0 \; ; \prec) \\
= \; &(\overline{\mathsf{rt}} \cup \mathsf{so} \cup \mathsf{ar}_0 \cup ((\mathsf{vis}_0 \setminus \mathsf{so}) \; ; \; \mathsf{rt}))^+ \cup (\overline{\mathsf{rt}} \cup \mathsf{so} \cup \mathsf{ar}_0 \cup ((\mathsf{vis}_0 \setminus \mathsf{so}) \; ; \; \mathsf{rt}))^* \; ; \prec \\
\subseteq \; &(\overline{\mathsf{rt}} \cup \mathsf{so} \cup \mathsf{ar}_0 \cup ((\mathsf{vis}_0 \setminus \mathsf{so}) \; ; \; \mathsf{rt}))^+ \cup \prec \cup \, \mathsf{ar}_0 \; ; \prec. && \text{by (15)}
\end{aligned}$$

But this contradicts Lemma 29 and Proposition 33. ◻

▶ **Proposition 34.** $\prec$ *is prefix-finite.*

**Proof.** Fix $f \in E$. By Proposition 31 it is enough to show that there are only finitely many $e$ such that $e \prec f$. By Eventual, there are only finitely many $g$ such that $\mathsf{obj}(g) = \mathsf{obj}(f)$ and $\neg(f \xrightarrow{\mathsf{vis}} g)$. Since $\mathsf{so}_0$, $\mathsf{rt}$, $\mathsf{vis}_0$ and $\mathsf{ar}_0$ are prefix-finite, for a $g$ satisfying the above property there are only finitely many $e$ such that

$$(e, g) \in ((\mathsf{vis}_0 \setminus \mathsf{so}) \cup \langle \mathsf{EPush} \rangle) \; ; (\mathsf{rt} \cap (\mathsf{Event} \times \mathsf{EPull})) \; ; \mathsf{so}_0?$$

This implies the required. ◻

## APPENDIX

▶ **Proposition 35.** $\overline{\mathsf{rt}} \cup \mathsf{so} \cup \mathsf{ar}_0 \cup ((\mathsf{vis}_0 \setminus \mathsf{so}) \mathbin{;} \mathsf{rt}) \cup \prec$ *is prefix-finite.*

**Proof.** Follows from Propositions 28, 32 and 34. ◻

**Proof of Lemma 8.** If for no event $e$ we have

$$|\{f \in E \mid \neg(e \xrightarrow{\mathsf{vis}} f)\}| = \infty, \tag{16}$$

then we are done. Otherwise, choose as $e$ the minimal such event in $\mathsf{ar}$. The order $\mathsf{so}$ is a union of total orders over a finite number of disjoint subsets of $E$, representing the sessions of $\mathcal{A}$. Let $E_1, \dots, E_n$ be these subsets. We say that a session $E_i$ does not ignore $e$ if either $E_i$ is finite or for some $f \in E_i$ we have $e \xrightarrow{\mathsf{vis}} f$. For any such session, by MonotonicView we have

$$|\{f \in E_i \mid \neg(e \xrightarrow{\mathsf{vis}} f)\}| < \infty.$$

Let $E_{i_1}, \dots, E_{i_m}$ be the sessions that ignore $e$; by (16), there is at least one such session. Let $E'_{i_1}, \dots, E'_{i_m}$ be respective suffixes of $E_{i_1}, \dots, E_{i_m}$ in $\mathsf{so}$ such that:

1. $\forall f \in E.\, f \xrightarrow{\mathsf{ar}} e \implies \forall j. \forall g \in E'_{i_j}.\, f \xrightarrow{\mathsf{vis}} g$;
2. $\forall j. \forall f \in E'_{i_j}.\, \mathsf{obj}(f) \neq \mathsf{obj}(e)$;
3. $\forall j. \forall f \in E'_{i_j}.\, \neg(f \xrightarrow{\mathsf{rt?;ar}} e)$;
4. $\forall j, k. \forall f \in E'_{i_j}. \forall g \in E_{i_k}.\, f \xrightarrow{\mathsf{rt} \cap (\mathsf{Event} \times \mathsf{EPull})} g \implies g \in E'_{i_k}$.

Such suffixes exist because:

1. $e$ is the minimal event in $\mathsf{ar}$ for which (16) holds;
2. a session $E_{i_j}$ may not contain infinitely many events on $\mathsf{obj}(e)$: if this were the case, then by (2) the session could not ignore $e$.
3. $\mathsf{ar}$ and $\mathsf{rt}$ are prefix-finite;
4. $\mathsf{rt}$ is prefix-finite.

Let

$$\mathsf{vis}_1 = \mathsf{vis} \cup \{(e, f) \mid \exists j.\, f \in E'_{i_j}\}.$$

Then

$$|\{f \in E \mid \neg(e \xrightarrow{\mathsf{vis}_1} f)\}| < \infty.$$

Due to item 3 above, we have $\mathsf{vis}_1 \subseteq \mathsf{ar}$. The tuple $(\mathcal{H}, \mathsf{vis}_1, \mathsf{ar})$ satisfies RYW, PushedVis, PushedAr, because so does $(\mathcal{H}, \mathsf{vis}, \mathsf{ar})$. The tuple $(\mathcal{H}, \mathsf{vis}_1, \mathsf{ar})$ also satisfies the following axioms:

MonotonicView: by construction;
ObservedVis: due to items 1 and 4 above;
ObservedAr: due to item 3 above;
RetVal: due to item 2 above.

In particular, the above axioms imply that $\mathsf{vis}_1$ is transitive. Then it is easy to see that it is also prefix-finite.

Thus, $(\mathcal{H}, \mathsf{vis}_1, \mathsf{ar})$ is an execution that satisfies all GSP axioms except possibly Eventual, and where no session ignores $e$. Continuing the above process *ad infinitum*, we can construct $\mathsf{vis}' \supseteq \mathsf{vis}$ such that $(\mathcal{H}, \mathsf{vis}', \mathsf{ar})$ satisfies all GSP axioms including Eventual. Using item 3 above, it is also easy to check that $\mathsf{vis}'$ constructed in this way is prefix-finite, so $(\mathcal{H}, \mathsf{vis}', \mathsf{ar})$ is the desired execution. ◻

## APPENDIX

**Proof of Lemma 5.** It is easy to see that $\mathsf{vis}_0 \subseteq \mathsf{vis}$ and $\mathcal{A} = (\mathcal{H}, \mathsf{vis}, \mathsf{ar})$ satisfies RYW, MonotonicView and PushedVis. $\mathcal{A}$ satisfies ObservedVis, because

$$\begin{aligned}
&\mathsf{ar}? \,;\, (\mathsf{vis} \setminus \mathsf{so}) \,;\, (\mathsf{rt} \cap (\mathsf{Event} \times \mathsf{EPull}))? \\
\subseteq\ &\mathsf{ar}? \,;\, ((\mathsf{ar}? \,;\, (\mathsf{vis}_0 \setminus \mathsf{so}) \,;\, (\mathsf{rt} \cap (\mathsf{Event} \times \mathsf{EPull}))? \,;\, \mathsf{so}?) \cup \\
&\qquad ((\mathsf{ar}? \,;\, (\mathsf{rt}? \cap (\mathsf{EPush} \times \mathsf{EPull})) \,;\, \mathsf{so}?) \setminus \mathsf{Id}) \,;\, (\mathsf{rt} \cap (\mathsf{Event} \times \mathsf{EPull}))? \\
\subseteq\ &(\mathsf{ar}? \,;\, (\mathsf{vis}_0 \setminus \mathsf{so}) \,;\, (\mathsf{rt} \cap (\mathsf{Event} \times \mathsf{EPull}))? \,;\, \mathsf{so}?) \cup ((\mathsf{ar}? \,;\, (\mathsf{rt}? \cap (\mathsf{EPush} \times \mathsf{EPull})) \,;\, \mathsf{so}?) \setminus \mathsf{Id}) \\
\subseteq\ &\mathsf{vis}.
\end{aligned}$$

Hence, $\mathsf{vis}_0 \subseteq \mathsf{vis}$ and $\mathcal{A} = (\mathcal{H}, \mathsf{vis}, \mathsf{ar})$ satisfies RYW-ObservedAr. We now show that $\mathsf{vis}$ is the minimal relation with this property. Consider another such relation $\mathsf{vis}'$. Then $\mathsf{vis}_0 \subseteq \mathsf{vis}'$ and $\mathcal{A}' = (\mathcal{H}, \mathsf{vis}', \mathsf{ar})$ satisfies RYW-ObservedAr. By RYW we have $\mathsf{so} \subseteq \mathsf{vis}'$. Since $\mathsf{vis}_0 \subseteq \mathsf{vis}'$, by ObservedVis and MonotonicView we have

$$\mathsf{ar}? \,;\, (\mathsf{vis}_0 \setminus \mathsf{so}) \,;\, (\mathsf{rt} \cap (\mathsf{Event} \times \mathsf{EPull}))? \,;\, \mathsf{so}? \subseteq \mathsf{vis}'.$$

Finally, by PushedVis and MonotonicView we have

$$(\mathsf{ar}? \,;\, (\mathsf{rt}? \cap (\mathsf{EPush} \times \mathsf{EPull})) \,;\, \mathsf{so}?) \setminus \mathsf{Id} \subseteq \mathsf{vis}'.$$

Hence, $\mathsf{vis} \subseteq \mathsf{vis}'$. ◻

**Proof of Theorem 2.** Let $\mathsf{ar}$ be any total prefix-finite order containing $\overline{\mathsf{rt}} \cup \mathsf{so} \cup \mathsf{ar}_0 \cup ((\mathsf{vis}_0 \setminus \mathsf{so}) \,;\, \mathsf{rt}) \cup \prec$; such an order exists by Lemma 6, Proposition 35 and Lemma 20. We let $\mathsf{vis}$ be defined as in Lemma 5 and let $\mathcal{A} = (\mathcal{H}, \mathsf{vis}, \mathsf{ar})$. Then the tuple $\mathcal{A}$ satisfies RYW-PushedVis. It satisfies PushedAr because $\overline{\mathsf{rt}} \subseteq \mathsf{ar}$. It also satisfies ObservedAr, because

$$\begin{aligned}
&(\mathsf{vis} \setminus \mathsf{so}) \,;\, \mathsf{rt} \\
\subseteq\ &(\mathsf{ar}? \,;\, (\mathsf{vis}_0 \setminus \mathsf{so}) \,;\, (\mathsf{rt} \cap (\mathsf{Event} \times \mathsf{EPull}))? \,;\, \mathsf{so}?) \,;\, \mathsf{rt} \cup (\mathsf{ar}? \,;\, (\mathsf{rt}? \cap (\mathsf{EPush} \times \mathsf{EPull})) \,;\, \mathsf{so}?) \,;\, \mathsf{rt} \\
\subseteq\ &(\mathsf{ar}? \,;\, (\mathsf{vis}_0 \setminus \mathsf{so}) \,;\, \mathsf{rt}) \cup (\mathsf{ar}? \,;\, (\mathsf{rt} \cap (\mathsf{EPush} \times \mathsf{Event}))) \\
\subseteq\ &(\mathsf{ar}? \,;\, \mathsf{ar}) \cup (\mathsf{ar}? \,;\, \mathsf{ar}) \\
\subseteq\ &\mathsf{ar}.
\end{aligned}$$

It is easy to see that $\mathsf{vis} \subseteq \mathsf{ar}$. By Proposition 4, $\mathsf{vis}$ is transitive and, hence, prefix-finite. Thus, the tuple $\mathcal{A}$ is indeed an abstract execution.

We next argue that $\mathcal{A}$ satisfies RetVal, which exploits the particular way in which we constructed $\mathsf{ar}$. To this end, we show that for any object $x$ we have $\mathsf{vis}|_x = \mathsf{vis}_x$, where $\mathsf{vis}|_x$ is the projection of $\mathsf{vis}$ to events on $x$. Then since for any $x$ we have $\mathsf{ar}_x \subseteq \mathsf{ar}$ and $\mathcal{A}_x$ satisfies RetVal, so does $\mathcal{A}$.

Since $\mathsf{vis}_x \subseteq \mathsf{vis}$ by construction, we only need to show $\mathsf{vis}|_x \subseteq \mathsf{vis}_x$. Consider arbitrary $f, g \in E$ such that $\mathsf{obj}(f) = \mathsf{obj}(g) = x$ and $f \xrightarrow{\mathsf{vis}} g$. We have:

$$(f, g) \in \mathsf{so} \cup (\mathsf{ar}? \,;\, (\mathsf{vis}_0 \setminus \mathsf{so}) \,;\, (\mathsf{rt} \cap (\mathsf{Event} \times \mathsf{EPull}))? \,;\, \mathsf{so}?) \cup ((\mathsf{ar}? \,;\, (\mathsf{rt}? \cap (\mathsf{EPush} \times \mathsf{EPull})) \,;\, \mathsf{so}?) \setminus \mathsf{Id}).$$

To show $f \xrightarrow{\mathsf{vis}_x} g$, our proof considers several cases corresponding to which of the components of the above union the edge $(f, g)$ belongs to.

- $(f, g) \in \mathsf{so}$. Since $\mathsf{obj}(f) = \mathsf{obj}(g) = x$, by RYW for $\mathcal{A}_x$ we get $(f, g) \in \mathsf{vis}_x$.
- $(f, g) \in \mathsf{ar}? \,;\, (\mathsf{vis}_0 \setminus \mathsf{so}) \,;\, (\mathsf{rt} \cap (\mathsf{Event} \times \mathsf{EPull})) \,;\, \mathsf{so}?$. Then for some $g'$ we have

$$f \xrightarrow{\mathsf{ar}?;(\mathsf{vis}_0 \setminus \mathsf{so});(\mathsf{rt} \cap (\mathsf{Event} \times \mathsf{EPull}))} g' \xrightarrow{\mathsf{so}?} g.$$

If $\mathsf{obj}(g') \neq \mathsf{obj}(g)$, then since the history $\mathcal{H}$ is well-fenced, for some $g'' \in \mathsf{EPull}$ we have

$$g' \xrightarrow{\mathsf{so}} g'' \xrightarrow{\mathsf{so}_0?} g.$$

Since $\mathsf{so} \subseteq \mathsf{rt}$, this implies

$$g' \xrightarrow{\mathsf{rt} \cap (\mathsf{Event} \times \mathsf{EPull})} g'' \xrightarrow{\mathsf{so}_0?} g.$$

Hence,

$$f \xrightarrow{\mathsf{ar}?;(\mathsf{vis}_0 \setminus \mathsf{so});(\mathsf{rt} \cap (\mathsf{Event} \times \mathsf{EPull}))} g'' \xrightarrow{\mathsf{so}_0?} g. \tag{17}$$

If $\mathsf{obj}(g') = \mathsf{obj}(g)$, then $g' \xrightarrow{\mathsf{so}_0?} g$ and we again have (17) for $g'' = g'$. Thus, in all cases (17) holds for some $g''$.
Then for some $e$ we have

$$f \xrightarrow{\mathsf{ar}?} e \xrightarrow{(\mathsf{vis}_0 \setminus \mathsf{so});(\mathsf{rt} \cap (\mathsf{Event} \times \mathsf{EPull}))} g'' \xrightarrow{\mathsf{so}_0?} g.$$

Now if $\neg(f \xrightarrow{\mathsf{vis}_x} g)$, then $e \prec f$, contradicting the fact that $\prec \subseteq \mathsf{ar}$. Hence, we must have $f \xrightarrow{\mathsf{vis}_x} g$, as required.

- $(f, g) \in \mathsf{ar}?\ ;\ ((\mathsf{vis}_0 \setminus \mathsf{vis}_x) \setminus \mathsf{so})\ ;\ \mathsf{so}?$. Then for some $g'$ we have

$$f \xrightarrow{\mathsf{ar}?;(\mathsf{vis}_0 \setminus \mathsf{so})} g' \xrightarrow{\mathsf{so}?} g.$$

and $\mathsf{obj}(g') \neq \mathsf{obj}(g)$. Since the history $\mathcal{H}$ is well-fenced, for some $g'' \in \mathsf{EPull}$ we have

$$g' \xrightarrow{\mathsf{so}} g'' \xrightarrow{\mathsf{so}_0?} g,$$

so that

$$f \xrightarrow{\mathsf{ar}?;(\mathsf{vis}_0 \setminus \mathsf{so});(\mathsf{rt} \cap (\mathsf{Event} \times \mathsf{EPull}))} g'' \xrightarrow{\mathsf{so}_0?} g.$$

We thus end up with the previous case.

- $(f, g) \in \mathsf{ar}?\ ;\ (\mathsf{vis}_x \setminus \mathsf{so})\ ;\ \mathsf{so}?$. Since $\mathsf{obj}(f) = \mathsf{obj}(g) = x$, this implies $(f, g) \in \mathsf{ar}_x?\ ;\ (\mathsf{vis}_x \setminus \mathsf{so})\ ;\ \mathsf{so}_0?$. Then by ObservedVis and MonotonicView for $\mathcal{A}_x$ we get $f \xrightarrow{\mathsf{vis}_x} g$.
- $(f, g) \in (\mathsf{ar} \cap (\mathsf{Event} \times (\mathsf{EPush} \cap \mathsf{EPull})))\ ;\ \mathsf{so}_0?$. Since $\mathsf{obj}(f) = \mathsf{obj}(g) = x$, this implies $(f, g) \in (\mathsf{ar}_x \cap (\mathsf{Event} \times (\mathsf{EPush} \cap \mathsf{EPull})))\ ;\ \mathsf{so}_0?$. Then by PushedVis and MonotonicView for $\mathcal{A}_x$ we get $f \xrightarrow{\mathsf{vis}_x} g$.
- $(f, g) \in \mathsf{ar}?\ ;\ (\mathsf{rt} \cap (\mathsf{EPush} \times \mathsf{EPull}))\ ;\ \mathsf{so}_0?$. Then for some $g'$ and $e$ we have

$$f \xrightarrow{\mathsf{ar}?} e \xrightarrow{(\mathsf{rt} \cap (\mathsf{EPush} \times \mathsf{EPull}))} g' \xrightarrow{\mathsf{so}_0?} g.$$

Now if $\neg(f \xrightarrow{\mathsf{vis}_x} g)$, then $e \prec f$, contradicting the definition of $\mathsf{ar}$. Hence, we must have $f \xrightarrow{\mathsf{vis}_x} g$.

- $(f, g) \in \mathsf{ar}?\ ;\ (\mathsf{rt}? \cap (\mathsf{EPush} \times \mathsf{EPull}))\ ;\ (\mathsf{so} \setminus \mathsf{so}_0)$. Then for some $g'$ we have

$$f \xrightarrow{\mathsf{ar}?;(\mathsf{rt}? \cap (\mathsf{EPush} \times \mathsf{EPull}))} g' \xrightarrow{\mathsf{so}} g$$

and $\mathsf{obj}(g') \neq \mathsf{obj}(g)$. Since the history $\mathcal{H}$ is well-fenced, for some $g'' \in \mathsf{EPull}$ we have

$$g' \xrightarrow{\mathsf{so}} g'' \xrightarrow{\mathsf{so}_0?} g,$$

so that

$$f \xrightarrow{\mathsf{ar}?;(\mathsf{rt} \cap (\mathsf{EPush} \times \mathsf{EPull}))} g'' \xrightarrow{\mathsf{so}_0?} g.$$

We thus end up with the previous case.

Thus, in all cases we have $f \xrightarrow{\mathsf{vis}_x} g$, validating RetVal.

Hence, $\mathcal{A}$ satisfies all GSP axioms except for Eventual. Since $\forall x.\, \mathsf{vis}|_x = \mathsf{vis}_x$ and $\mathcal{A}_x$ satisfies Eventual, we have

$$\forall e \in E.\, |\{f \in E \mid \mathsf{obj}(e) = \mathsf{obj}(f) \wedge \neg(e \xrightarrow{\mathsf{vis}} f)\}| < \infty.$$

Then by Lemma 8 there exists $\mathsf{vis}'$ such that $(\mathcal{H}, \mathsf{vis}', \mathsf{ar}) \in \mathsf{ExecGSC}$. □